\documentstyle[a4,11pt]{article}

\begin{document}

\title{{\bf Topological Extensions of Noether Charge Algebras carried by D-}$p${\bf %
-branes} \\
[1cm]}
\author{Hanno Hammer \\
DAMTP, University of Cambridge,\\
Silver St., Cambridge, UK\\
[1cm]}
\date{2.11.1997 \\
[1cm]}
\maketitle

\begin{abstract}
We derive the fully extended supersymmetry algebra carried by D-branes in a
massless type IIA superspace vacuum. We find that the extended algebra
contains not only topological charges that probe the presence of compact
spacetime dimensions but also pieces that measure non-trivial configurations
of the gauge field on the worldvolume of the brane. Furthermore there are
terms that measure the coupling of the non-triviality of the worldvolume
regarded as a U(1)-bundle of the gauge field to possible compact spacetime
dimensions. In particular, the extended algebra carried by the D-2-brane can
contain the charge of a Dirac monopole of the gauge field. In the course of
this work we derive a set of generalized Gamma-matrix identities that
include the ones presently known for the IIA case. -- In the first part of
the paper we give an introduction to the basic notions of Noether current
algebras and charge algebras; furthermore we find a Theorem that describes
in a general context how the presence of a gauge field on the worldvolume of
an embedded object transforming under the symmetry group on the target space
alters the algebra of the Noether charges, which otherwise would be the same
as the algebra of the symmetry group. This is a phenomenon recently found by
Sorokin and Townsend in the case of the M-5-brane, but here we show that it
holds quite generally, and in particular also in the case of D-branes.
\end{abstract}

\newpage

\tableofcontents

\newpage

\section*{Introduction}

Topological extensions of the algebra of Noether currents and corresponding
Noether charges have been studied in the past by a number of authors \cite
{Azca,Soro}. In \cite{Azca} the extensions of the algebra of Noether
and modified Noether charges carried by supersymmetric extended objects have
been examined; furthermore, it has been pointed out that the origin of these
modifications is the Wess-Zumino term in the Lagrangian of the extended
object. In \cite{Soro}
the algebra of the Noether supercharges of the M-$5$-brane was derived, and
it was observed that not all central charges occuring in the superalgebra
extension are entirely due to the Wess-Zumino term; it was shown that
another contribution to the central charges originates in the presence of a
gauge field potential on the worldvolume of the M-$5$-brane, which takes
part in the action of the superPoincare group acting on the target space. In
this work we shall prove a theorem explaining that this is a general feature
for a whole class of Lagrangians containing a gauge potential on the
worldvolume which is forced to transform under a Lie group that acts on the
target space of the theory; in this case even the algebra of the
(unmodified) Noether charges suffers modifications, which otherwise would
close into the original algebra of the group that acts on the target space,
possibly up to a sign, which is a consequence of whether the group acts from
the left or from the right.

In this work we have performed an analysis of the extensions of the
superalgebra of Noether and modified Noether charges carried by D-$p$-branes
in a IIA superspace. In doing so, however, we have faced a number of
difficulties which could not be illuminated by consulting the literature; in
section \ref{NoethChar} we therefore provide an introduction to the basic
concepts of the algebra of Noether currents, Noether charges, and associated
modified currents and charges, which might arise from a Lagrangian
transforming as a semi-invariant under the action of the group. We show that
for a left action the algebra of the associated Noether charges always
closes to the original algebra, regardless of whether the Lagrangian is
invariant under the transformation or not; we derive the action of the
Noether charges on the currents and show that for a left action the Noether
currents transfom in the adjoint representation of the group. We then extend
this analysis to the case of semi-invariant Lagrangians and examine
carefully under which circumstances certain contributions to the Poisson
brackets of the modified currents vanish or may be neglected; this question
is not always fully adressed in the literature, and becomes even more
non-trivial in the case of having a gauge field present on the worldvolume,
since the gauge field degrees of freedom are subject to primary and
secondary constraints (in Dirac's terminology). We analyze the constraint
structure of a theory possessing such a gauge field on the world-volume; the
results apply to D-branes and M-branes as well. We derive conditions under
which the additional charges obtained so far are conserved and central. We
finish the first section with showing how modifications of the Noether
charge algebra arise from the presence of such a gauge field, even if the
modifications of the charge algebra due to semi-invariant pieces in the
Lagrangian are not yet taken into account.

In section \ref{SuperAlgebra} we apply these ideas to derive the extensions
of the algebra of modified Noether charges for D-$p$-brane Lagrangians in
IIA superspace. We first derive a general form of these extensions
applicable for the most general forms of gauge fields (NS-NS and RR) on the
superspace; then we choose a particular background in putting all bosonic
components of the gauge fields to zero, and taking into account that the
remainder are subject to superspace constraints which allow to reconstruct
the leading components of the RR gauge field strengths unambiguously. In
doing so we must check whether the RR field strengths thus derived actually
satisfy the appropriate Bianchi identities; we find that this question can
be traced back to the validity of a set of generalized $\Gamma $-matrix
identities; it is known that the first two members in this set are actually
valid; as for the rest we derive a necessary condition using the technique
that has been applied in similar circumstances previously, see \cite
{Achu,PKT1}, and find that it is satisfied. The extended algebra thus
derived contains topological charges that probe the existence of compact
spacetime dimensions the brane is wrapping around; furthermore, which is a
new feature here, we find that central charges show up that probe the
non-triviality of the worldvolume regarded as a $U\left( 1\right) $-bundle
of the gauge field $A_\mu $; we find that for the D-$4$-, $6$- and $8$%
-branes there exist central charges originating in the Wess-Zumino term that
can be interpreted as probing the coupling of these non-trivial gauge-field
configurations to compact dimensions in the spacetime; they are zero if
either there are no compact dimensions, or the brane is not wrapping around
them, or the $U\left( 1\right) $-bundle is trivial, which requires the gauge
field configuration to be trivial. As for the D-$2$-brane such a coupling of
spacetime topology to the gauge field is only present in the central charges
that stem from the fact that the gauge field transforms under supersymmetry;
they have nothing to do with the Wess-Zumino term; for the special case of a
D-$2$-brane given by ${\bf R}\times S^2$, where ${\bf R}$ denotes the time
dimension, we find that the algebra can contain the charge of a Dirac
monopole of the gauge field; this result is very neat, so we present it here:%
$$
\left\{ Q_\alpha ,Q_\beta \right\} =2\left( C\Gamma ^m\right) _{\alpha \beta
}\cdot P_m\;-\;2i\left( C\Gamma _{11}\Gamma _m\right) _{\alpha \beta }\cdot
Y^m\;- 
$$
$$
-\;i\left( C\Gamma _{m_2m_1}\right) _{\alpha \beta }\cdot
T^{m_1m_2}\;-\;2i\left( C\Gamma _{11}\right) _{\alpha \beta }\cdot 4\pi
g\quad . 
$$
Here $Y^m$ is a central charge that couples the canonical gauge field
momentum to compact dimensions in the spacetime allowing for $1$-cycles in
the brane wrapping around them; $T^{m_1m_2}$ probes the presence of compact
dimensions in spacetime the brane wraps around, i.e. allowing for $2$-cycles
wrapping around them, and $g$ is the quantized charge of a Dirac monopole
resulting from the gauge field.

\section{Topological extensions of the algebra of Noether charges}

\subsection{Actions of a Lie group and associated Noether charges\quad \quad 
\label{NoethChar}}

\subsubsection{Noether currents}

Let $\left( x^\mu \right) =\left( t,\sigma ^r\right) $, $\mu =0,\ldots ,p$; $%
r=1,\ldots ,p$ denote coordinates on a $\left( p+1\right) $-dimensional
manifold (''worldvolume'') $W$. Here $t$ refers to a ''timelike''
coordinate, $\sigma ^r$ refers to ''spacelike'' coordinates. Let $W\left(
t\right) $ denote the hypersurfaces in $W$ with constant $t$. Let ${\cal L}=%
{\cal L}\left( \phi ,\partial _\mu \phi \right) $ be a Lagrangian of a field
multiplet $\phi ={{\phi ^1} \choose {\vdots }}$ defined on $W$ , with unspecified
dimension. The objects $\left( \phi ^i\right) $ are regarded to be
coordinates on a target space $\Sigma $; at present we do not make any
further assumptions on the precise nature of $\Sigma $. Let $G$ be a Lie
group with generators $T_M\in Lie\left( G\right) $, where $Lie\left(
G\right) $ is the Lie algebra of $G$; the generators $T_M$ act on $\phi ^i$
according to $\phi \mapsto \delta _M\phi ={{\delta _M\phi ^1} \choose {\vdots }}$%
; here $\delta _M\phi ^i$ are the components of the vector field $\widetilde{%
T_M}$ induced by the generator $T_M$ on $\Sigma $, i.e., the action of $%
e^{t\cdot T_M}$ defines a flow $\left( \phi ,t\right) \mapsto \left(
e^{t\cdot T_M}\phi \right) ^i$, which generates the vector field \cite
{cram/pir} 
\begin{equation}
\label{gract0.0}\left. \frac d{dt}\left( e^{t\cdot T_M}\phi \right)
^i\right| _{t=0}\frac \partial {\partial \phi ^i}=\left. \left( \widetilde{%
T_M}\right) ^i\right| _\phi \frac \partial {\partial \phi ^i}=\left( \delta
_M\phi ^i\right) \frac \partial {\partial \phi ^i}=\widetilde{T_M}\quad . 
\end{equation}
For a right action the map $Lie\left( G\right) \ni X\mapsto \tilde X$, which
sends an element of the Lie algebra of $G$ to an induced vector field on $%
\Sigma $, is a Lie algebra homomorphism into the set of all vector fields on 
$\Sigma $ endowed with the Lie bracket as multiplication: 
\begin{equation}
\label{zw1}\widetilde{\left[ X,Y\right] }=\left[ \widetilde{X},\widetilde{Y}%
\right] \quad . 
\end{equation}
For a left action this is true for the map $Lie\left( G\right) \ni X\mapsto
-\tilde X$, since in this case 
\begin{equation}
\label{zw2}\widetilde{\left[ X,Y\right] }=-\left[ \widetilde{X},\widetilde{Y}%
\right] \quad . 
\end{equation}
Now denote the expression for the equations of motion for the fields $\phi
^i $ by 
\begin{equation}
\label{gract1}\left( eq,{\cal L}\right) _i:=\frac{\partial {\cal L}}{%
\partial \phi ^i}-\partial _\mu \frac{\partial {\cal L}}{\partial \partial
_\mu \phi ^i}\quad , 
\end{equation}
then the action of the generator $T_M$ on ${\cal L}$ takes the form 
\begin{equation}
\label{gract2}\delta _M{\cal L}=\delta _M\phi ^i\cdot \left( eq,{\cal L}%
\right) _i+\partial _\mu j_M^\mu \quad , 
\end{equation}
where 
\begin{equation}
\label{gract3}j_M^\mu =\delta _M\phi ^i\cdot \frac{\partial {\cal L}}{%
\partial \partial _\mu \phi ^i} 
\end{equation}
is the {\it Noether current associated with} $T_M$. If $sol\left( {\cal L}%
\right) $ denotes a solution to the {\it equations of motion} $\left( eq,%
{\cal L}\right) _i=0$, we have 
\begin{equation}
\label{gract4}\left[ \delta _M{\cal L}=\partial _\mu j_M^\mu \right]
_{sol\left( {\cal L}\right) }\quad , 
\end{equation}
i.e. on the solution $sol\left( {\cal L}\right) $. Now assume that ${\cal L}=%
{\cal L}_0+{\cal L}_1$, where ${\cal L}_0$ is invariant under $G$, $\delta _M%
{\cal L}_0=0$. Then 
\begin{equation}
\label{gract5}\delta _M{\cal L}_1=\delta _M\phi ^i\cdot \left( eq,{\cal L}_0+%
{\cal L}_1\right) _i+\partial _\mu j_M^\mu \quad , 
\end{equation}
where 
\begin{equation}
\label{gract6}j_M^\mu =\delta _M\phi ^i\cdot \frac{\partial {\cal L}_0}{%
\partial \partial _\mu \phi ^i}+\delta _M\phi ^i\cdot \frac{\partial {\cal L}%
_1}{\partial \partial _\mu \phi ^i}=:j_{0,M}^\mu +J_M^\mu \quad . 
\end{equation}
Therefore, 
\begin{equation}
\label{gract7}\delta _M{\cal L}_1=\delta _M\phi ^i\cdot \left( eq,{\cal L}_0+%
{\cal L}_1\right) _i+\partial _\mu \left( j_{0,M}^\mu +J_M^\mu \right) \quad
, 
\end{equation}
and 
\begin{equation}
\label{gract8}\left[ \delta _M{\cal L}_1=\partial _\mu \left( j_{0,M}^\mu
+J_M^\mu \right) \right] _{sol\left( {\cal L}_0+{\cal L}_1\right) }\quad . 
\end{equation}
This is to be compared with 
\begin{equation}
\label{gract9}0=\delta _M\phi ^i\cdot \left( eq,{\cal L}_0\right)
_i+\partial _\mu j_{0,M}^\mu \quad , 
\end{equation}
and 
\begin{equation}
\label{gract10}\left[ 0=\partial _\mu j_{0,M}^\mu \right] _{sol\left( {\cal L%
}_0\right) }\quad . 
\end{equation}
Since%
$$
\left[ \partial _\mu j_{0,M}^\mu \right] _{sol\left( {\cal L}_0+{\cal L}%
_1\right) }\neq \left[ \partial _\mu j_{0,M}^\mu \right] _{sol\left( {\cal L}%
_0\right) }=0 
$$
in general, we see that $j_{0,M}^\mu $ is no longer conserved in the
presence of ${\cal L}_1$, although it is conserved on the critical
trajectories of ${\cal L}_0$. Neither is the total current conserved,%
$$
\left[ \delta _M{\cal L}_1=\partial _\mu j_M^\mu \right] _{sol\left( {\cal L}%
_0+{\cal L}_1\right) }\quad . 
$$

To proceed, we now specify the action of $G$ on ${\cal L}_1$: We assume
that, under the action of $G$, ${\cal L}_1$ transforms as a total derivative 
{\bf on- and off-shell}, i.e. without using the equations of motion. Then $%
{\cal L}_1$ is said to be {\it semi-invariant} under the action of $G$. This
means that 
\begin{equation}
\label{gract11}\delta _M{\cal L}_1=\partial _\mu U_M^\mu \quad , 
\end{equation}
for some functions $U_M^\mu $ of the fields and its derivatives. This gives,
using (\ref{gract7}),%
$$
0=\delta _M\phi ^i\cdot \left( eq,{\cal L}_0+{\cal L}_1\right) _i+\partial
_\mu \left( j_M^\mu -U_M^\mu \right) \quad , 
$$
and we see that the modified current 
\begin{equation}
\label{gract12}\widetilde{j_M^\mu }:=j_M^\mu -U_M^\mu 
\end{equation}
is conserved on the critical trajectories of ${\cal L}_0+{\cal L}_1$, i.e. 
\begin{equation}
\label{gract12.1}\partial _\mu \widetilde{j_M^\mu }=0\quad . 
\end{equation}
Note that in this case the {\bf conserved current is no longer a Noether
current}.

\subsubsection{Algebra of Poisson brackets\quad \quad \label{PB Algebra}}

The Poisson brackets of the zero components $j_M^0$ of the {\bf total}
Noether currents $j_M^\mu $ associated with the action of $T_M$ on some
Lagrangian ${\cal L}$ satisfy the Lie algebra of $G$, possibly up to a sign,
regardless of whether ${\cal L}$ is invariant or not. This can be proven by
introducing canonical momenta%
$$
\Lambda _i:=\frac{\partial {\cal L}}{\partial \dot \phi ^i}\quad , 
$$
so that 
\begin{equation}
\label{gract13}j_M^0=\delta _M\phi ^i\cdot \Lambda _i\quad .
\end{equation}
Let $C_{MN}^{\;K}$ denote the structure constants of the Lie algebra $%
Lie\left( G\right) $ of $G$, i.e.%
$$
\left[ T_M,T_N\right] =C_{MN}^{\;K}\cdot T_K\quad , 
$$
where $\left[ \cdot ,\cdot \right] $ denotes a (graded) commutator. Working
out the Poisson bracket we get 
\begin{equation}
\label{gr01}\left\{ j_M^0\left( t,\sigma \right) ,j_N^0\left( t,\sigma
^{\prime }\right) \right\} _{PB}=-\delta _M\phi ^i\frac{\partial \delta
_N\phi ^j}{\partial \phi ^i}\Lambda _j\,\delta \left( \sigma -\sigma
^{\prime }\right) +\delta _N\phi ^j\frac{\partial \delta _M\phi ^i}{\partial
\phi ^j}\Lambda _i\,\delta \left( \sigma -\sigma ^{\prime }\right) \quad .
\end{equation}
where $\left\{ \cdot ,\cdot \right\} _{PB}$ denotes a (graded) Poisson
bracket with canonical variables $\phi ^i$, $\Lambda _i$. If we now use the
results from (\ref{gract0.0}) we find%
$$
\left( \ref{gr01}\right) =-\left[ \widetilde{T_M},\widetilde{T_N}\right]
^i\Lambda _i\,\delta \left( \sigma -\sigma ^{\prime }\right) \quad . 
$$
Taking account of (\ref{zw1}, \ref{zw2}) this gives 
\begin{equation}
\label{gra131}\left\{ j_M^0\left( t,\sigma \right) ,j_N^0\left( t,\sigma
^{\prime }\right) \right\} _{PB}\;=\;\pm C_{MN}^{\;K}\cdot j_K^0\left(
t,\sigma \right) \cdot \delta \left( \sigma -\sigma ^{\prime }\right) \quad ,
\end{equation}
where $"+/-"$ refers to a {\bf left / right} action. If we define associated 
{\it Noether charges} 
\begin{equation}
\label{gract25}Q_M\left( t\right) :=\int\limits_{W\left( t\right) }d^p\sigma
\cdot j_M^0\left( t,\sigma \right) \quad ,
\end{equation}
then the once integrated version of (\ref{gra131}) is 
\begin{equation}
\label{gra132}\left\{ Q_M,j_N^0\right\} _{PB}=\pm \;j_K^0\cdot ad\left(
T_M\right) _{\;N}^K\quad ,
\end{equation}
where $ad\left( T\right) $ denotes the adjoint representation of the Lie
algebra element $T$. This now means that the total Noether currents span the
adjoint representation of $G$ in the case of a left action. Moreover, in
this case a further integration of (\ref{gra132}) yields back the algebra we
have started with, 
\begin{equation}
\label{gra133}\left\{ Q_M,Q_N\right\} _{PB}=\pm C_{MN}^{\;K}\cdot Q_K\quad .
\end{equation}

We omit the subscript $PB$ in what follows, and reintroduce it only when
there is danger of confusion with an anticommutator.

Now we look at the situation when the Lagrangian contains a semi-invariant
piece ${\cal L}_1$. In this case the conserved currents are $\widetilde{%
j_M^\mu }=j_M^\mu -U_M^\mu $, and their zero components have Poisson bracket
relations 
\begin{equation}
\label{gract15}\left\{ \widetilde{j_M^0}\left( t,\sigma \right) ,\widetilde{%
j_N^0}\left( t,\sigma ^{\prime }\right) \right\} =\left\{
j_M^0,j_N^0\right\} -\left\{ j_M^0,U_N^0\right\} -\left\{
U_M^0,j_N^0\right\} +\left\{ U_M^0,U_N^0\right\} \quad . 
\end{equation}
The brackets $\left\{ j_M^0,U_N^0\right\} =\left\{ \delta _M\phi ^i\Lambda
_i,U_N^0\right\} $ are always unequal zero when $U_N^0$ is not a constant,
since the presence of the canonical momenta amounts to derivatives with
respect to the fields on $U_N^0$. However, the brackets $\left\{
U_M^0,U_N^0\right\} $ are also non-vanishing in general; although in the
Lagrangian description they contained only fields $\phi ^i$ and their
derivatives, the shift to the Hamiltonian (first order) picture amounts to
inverting the relations%
$$
\Lambda _i=\frac{\partial {\cal L}}{\partial \dot \phi ^i}\left( \phi
,\partial _0\phi ,\partial _r\phi \right) 
$$
for $\partial _0\phi ^i$, which gives $\dot \phi ^i=\Phi ^i\left( \phi
,\partial _r\phi ,\Lambda \right) $, where $r,s=1,\ldots ,p$ refers to the
''spatial'' coordinates on $W$. Therefore,%
$$
U_M^0\left( \phi ,\partial _\mu \phi \right) \longrightarrow U_M^0\left(
\phi ,\Phi \left( \phi ,\partial _r\phi ,\Lambda \right) ,\partial _s\phi
\right) =\widehat{U_M^0}\left( \phi ,\partial _r\phi ,\Lambda \right) \quad
, 
$$
so that {\bf after Legendre transforming} the $U_M^0$ {\bf do} depend on the
canonical momenta, which makes their mutual Poisson brackets in general
non-vanishing. Note that the ''hatted'' $\widehat{U_M^0}$ is of course a
different function of its arguments than $U_M^0$ which makes itself manifest
when we are performing partial or functional derivatives, respectively.
Therefore, in a Poisson bracket we are always dealing with $\widehat{U_M^0}$%
; outside a Poisson bracket we can replace $\widehat{U_M^0}$ by $U_M^0$, as
we shall do in the following.

Now we subtract and add $\pm C_{MN}^{\;K}\cdot U_K^0\left( t,\sigma \right)
\cdot \delta \left( \sigma -\sigma ^{\prime }\right) $ on the right hand
side of (\ref{gract15}), and use (\ref{gra131}). This gives us 
\begin{equation}
\label{gract15.1}\left\{ \widetilde{j_M^0}\left( t,\sigma \right) ,%
\widetilde{j_N^0}\left( t,\sigma ^{\prime }\right) \right\} =\pm
C_{MN}^{\;K}\cdot \widetilde{j_K^0}\left( t,\sigma \right) \cdot \delta
\left( \sigma -\sigma \right) \;+\;\widetilde{S_{MN}^0}\;+\left\{ \widehat{%
U_M^0}\left( t,\sigma \right) ,\widehat{U_N^0}\left( t,\sigma ^{\prime
}\right) \right\} \quad , 
\end{equation}
with the ''anomalous'' piece%
$$
\widetilde{S_{MN}^0}+\left\{ \widehat{U_M^0}\left( t,\sigma \right) ,%
\widehat{U_N^0}\left( t,\sigma ^{\prime }\right) \right\} =-\left\{
j_M^0\left( t,\sigma \right) ,\widehat{U_N^0}\left( t,\sigma ^{\prime
}\right) \right\} -\left\{ \widehat{U_M^0}\left( t,\sigma \right)
,j_N^0\left( t,\sigma ^{\prime }\right) \right\} \;\pm 
$$
\begin{equation}
\label{gract16}\pm \;C_{MN}^{\;K}\cdot U_K^0\left( t,\sigma \right) \cdot
\delta \left( \sigma -\sigma ^{\prime }\right) \;+\left\{ \widehat{U_M^0}%
\left( t,\sigma \right) ,\widehat{U_N^0}\left( t,\sigma ^{\prime }\right)
\right\} . 
\end{equation}

Let us compute the brackets $\left\{ j_M^0,\widehat{U_N^0}\right\} $ for the
special case that the action of $G$ on covectors $\Lambda _i$ is specified
so as to make expressions like $\dot \phi ^i\Lambda _i$ transforming as
scalars under the group operation; this means that $\Lambda _i$ transform
contragrediently to $\phi ^i$, 
\begin{equation}
\label{gract17}\delta _M\Lambda _i=-\Lambda _j\frac{\partial \delta _M\phi ^j%
}{\partial \phi ^i}\quad .
\end{equation}
To see that this specification leaves $\dot \phi ^i\Lambda _i$ invariant we
apply $\delta _M$,%
$$
\delta _M\left( \dot \phi ^i\Lambda _i\right) =\left( \delta _M\frac
d{dt}\phi ^i-\dot \phi ^j\frac{\partial \delta _M\phi ^i}{\partial \phi ^j}%
\right) \Lambda _i\quad ; 
$$
if we assume now, as usual, that $\delta _M$ commutes with $\partial _\mu $
the expression in the bracket vanishes. We find 
\begin{equation}
\label{gr6}\left\{ j_M^0\left( t,\sigma \right) ,\widehat{U_N^0}\left(
t,\sigma ^{\prime }\right) \right\} =-\delta _MU_N^0\cdot \delta \left(
\sigma ^{\prime }-\sigma \right) \;+\frac \partial {\partial \sigma
^r}\left[ \delta _M\phi ^i\frac{\partial \widehat{U_N^0}}{\partial \partial
_r\phi ^i}\cdot \delta \left( \sigma ^{\prime }-\sigma \right) \right] \quad
.
\end{equation}
Double integration of the second term over $W\left( t\right) $, $t=const$.,
yields 
\begin{equation}
\label{gr3}\int\limits_{W\left( t\right) }d^p\sigma \cdot \frac \partial
{\partial \sigma ^r}\left[ \delta _M\phi ^i\frac{\partial \widehat{U_N^0}}{%
\partial \partial _r\phi ^i}\right] =\int\limits_{\partial W\left( t\right)
}d{\cal A}_r^{p-1}\cdot \delta _M\phi ^i\frac{\partial \widehat{U_N^0}}{%
\partial \partial _r\phi ^i}\quad ,
\end{equation}
where $d{\cal A}_r^{p-1}$ is a $\left( p-1\right) $-dimensional area
element. We must deal with this surface term appropriately. The manifold $%
W\left( t\right) $ can be infinitely extended in all spatial directions, or
some of these spatial directions may be compact. To avoid bothering with the
surface terms we assume from now on that the integrands of surface
contributions vanish sufficiently strong at the boundary $\partial W\left(
t\right) $, i.e. at points which lie at infinite values of the non-compact
coordinates. As a special case this includes the possibility that $W\left(
t\right) $ is closed, which implies that {\bf all} spatial coordinates $%
\sigma ^{\mu \,}$ are compact.Furthermore we assume that all expressions in
a total derivative, such as on the left hand side of (\ref{gr3}), are smooth
and defined {\bf globally} on $W$. (The emphasis on being globally defined
is of course to prevent us from situations where Stokes' theorem is not
applicable, i.e. ''surface terms cannot be integrated away''; this can be
true for the topological current to be defined below). Under these
circumstances {\bf all surface terms vanish}, and we obtain for the current
algebra 
\begin{equation}
\label{al}\left\{ \widetilde{j_M^0}\left( t,\sigma \right) ,\widetilde{j_N^0}%
\left( t,\sigma ^{\prime }\right) \right\} =\pm C_{MN}^{\;K}\cdot \widetilde{%
j_K^0}\left( t,\sigma \right) \cdot \delta \left( \sigma -\sigma ^{\prime
}\right) \;+\;\widetilde{S_{MN}^0}\;+\left\{ \widehat{U_M^0}\left( t,\sigma
\right) ,\widehat{U_N^0}\left( t,\sigma ^{\prime }\right) \right\} \quad ,
\end{equation}
\begin{equation}
\label{gract23}\widetilde{S_{MN}^0}\left( t,\sigma ,\sigma ^{\prime }\right)
=\left[ \delta _MU_N^0-\delta _NU_M^0\pm C_{MN}^{\;K}\cdot U_K^0\right]
\cdot \delta \left( \sigma -\sigma ^{\prime }\right) \quad +\quad \left( 
\mbox{total derivatives}\right) \quad ,
\end{equation}
with the total derivatives from (\ref{gr6}). Let us now define 
\begin{equation}
\label{gract251}Q_M\left( t\right) :=\int\limits_{W\left( t\right) }d^p\sigma
\cdot \widetilde{j_M^0}\left( t,\sigma \right) \quad ,
\end{equation}
\begin{equation}
\label{gract24}S_{MN}^\mu \left( t,\sigma \right) =\delta _MU_N^\mu -\delta
_NU_M^\mu \pm C_{MN}^{\;K}\cdot U_K^\mu \quad ,
\end{equation}
\begin{equation}
\label{gract26}Z_{MN}\left( t\right) :=\int\limits_{W\left( t\right)
}d^p\sigma \cdot S_{MN}^0\left( t,\sigma \right) \quad .
\end{equation}
Note that the charge $Q_M\left( t\right) =Q_M$ is no longer a Noether
charge, since it is defined through the conserved current $\widetilde{j_M^0}$
rather than the Noether current $j_M^0$. It is conserved, however, due to (%
\ref{gract12.1}). We show now that $Z_{MN}$ is conserved as well.

\subsubsection{Conservation of the new charges}

To prove this, observe that $\delta _M$ commutes with $\partial _\mu $;
therefore we can write%
$$
\partial _\mu S_{MN}^\mu =\delta _M\partial _\mu U_N^\mu -\delta _N\partial
_\mu U_M^\mu \pm C_{MN}^{\;K}\cdot \partial _\mu U_K^\mu \quad = 
$$
\begin{equation}
\label{gract27}=\quad \left( \delta _M\delta _N-\delta _N\delta _M\pm
C_{MN}^{\;K}\cdot \delta _K\right) {\cal L}_1\quad . 
\end{equation}
If we work out the double variation we find that the last expression
vanishes due to%
$$
\left[ \delta _M,\delta _N\right] {\cal L}=\left[ \delta _M,\delta _N\right]
\phi ^i\cdot {\cal L}_{\phi ^i}+\partial _\mu \left[ \delta _M,\delta
_N\right] \phi ^i\cdot {\cal L}_{\partial _\mu \phi ^i}\quad . 
$$
This can be seen yet in another way: On account of $\left[ \partial _\mu
,\delta _M\right] =0$, $\delta _N=\widetilde{T_N}$ acts on coordinates $\phi
^i$ in the same way as it acts on $\partial _\mu \phi ^i$. Therefore we can
replace the $\delta ^{\prime }$s in the round bracket in (\ref{gract27}) by
vector fields $\widetilde{T_N}$, which yields%
$$
\delta _M\delta _N-\delta _N\delta _M\pm C_{MN}^{\;K}\cdot \delta _K=\left[ 
\widetilde{T_M},\widetilde{T_N}\right] \pm C_{MN}^{\;K}\cdot \widetilde{T_K}%
\;= 
$$
$$
=\;\mp \widetilde{\left( \left[ T_M,T_N\right] -C_{MN}^{\;K}\cdot T_K\right) 
}=0\quad , 
$$
according to the algebra of the generators $\left( T_M\right) $. What we
have shown is 
\begin{equation}
\label{gract28}\partial _\mu S_{MN}^\mu =0\quad , 
\end{equation}
which is the local conservation law for the charge $Z_{MN}\left( t\right) $
defined in (\ref{gract26}).

Using the definitions (\ref{gract251}, \ref{gract26}), we find on double
integration of (\ref{al}) (and on assumption that this integration is
defined) 
\begin{equation}
\label{gract29}\left\{ Q_M,Q_N\right\} _{PB}=\pm C_{MN}^{\;K}\cdot
Q_K\;+\;Z_{MN}\;+\int\limits_{W\left( t\right) }d^p\sigma \,d^p\sigma
^{\prime }\cdot \left\{ \widehat{U_M^0}\left( t,\sigma \right) ,\widehat{%
U_N^0}\left( t,\sigma ^{\prime }\right) \right\} \quad .
\end{equation}
We see that our original algebra has been extended by conserved charges $%
Z_{MN}$; however, unless the Poisson brackets $\left\{ \widehat{U_M^0},%
\widehat{U_N^0}\right\} $ vanish, this extension does not close to a new
algebra!

\subsection{Coset spaces of Lie groups as target spaces\quad \quad \label
{CosetSpaces}}

\subsubsection{Closure of the algebra extension\quad \quad \label{Closure}}

In order to proceed further we now make more detailed assumptions about the
structure of the target space and the geometric origin of the invariant and
semi-invariant pieces in the Lagrangian. We assume that the target space is
now the group $G$ itself, with coordinates $\phi ^i$. More generally, we
could have that $G$ is a subgroup of a larger group $\tilde G$, which
contains yet another subgroup $H$ : $G,H\subset \tilde G$. Then $\Sigma $
could be the coset space $\tilde G/H$, and $G$ would act on elements of $%
\Sigma =\tilde G/H$ by left or right multiplication. This is the situation
we shall consider later, where $\tilde G=\,${\rm super}$Poincare$ in $D=10$
spacetime dimensions, $G$ is the subgroup generated by $\left\{ P_m,Q_\alpha
\right\} $, i.e. the generators of Poincare- and super-translations, and $H$
is the subgroup $SO\left( 1,9\right) $. If the objects $Q_\alpha $ build two 
$16$-component spinors with opposite chirality, then the coset space $\Sigma 
$ is type IIA superspace. However, for the purpose of illustrating of how
topological currents emerge we shall in the following refrain from any
graded groups, algebras, or whatsoever, and restrict ourselves to the
simpler case of $\Sigma =G$.

The fields $\phi ^i$ on $W$ accomplish an embedding $emb:W\rightarrow \Sigma 
$ of $W$ into $\Sigma $ by $emb\left( x\right) =\left( \phi ^1\left(
x\right) ,\ldots ,\phi ^{\dim G}\left( x\right) \right) $. From now on we
call $W$ the ''worldvolume'', following standard conventions. If the
hypersurfaces $W\left( t\right) $ are closed then the same holds for their
images in $\Sigma $, since $\partial \left[ embW\left( t\right) \right]
=emb\left[ \partial W\left( t\right) \right] =\emptyset $. In other words,
the images $embW\left( t\right) $ are $p${\it -cycles}$\ $ in $\Sigma $ in
this case. We assume that the previously made assumptions concerning surface
terms in integrands still hold, and that those spatial dimensions of $%
W\left( t\right) $ which are not infinitely extended are closed. Furthermore
we {\bf assume} that the semi-invariant piece ${\cal L}_1$ or Wess-Zumino
(WZ) term, as it will be called in the sequel, is the result of the
pull-back of a target space $\left( p+1\right) $-form $\left( WZ\right) $ to
the worldvolume $W$; from now on, we write ${\cal L}_1=:{\cal L}_{WZ}$ for
the semi-invariant piece. Its construction proceeds as follows:

Let $\left( \Pi ^A\right) _{A=1,\ldots ,\dim G}$ be left-invariant (LI) $1$%
-forms on $\Sigma =G$; this means, that at every point in $\Sigma $ they
span the cotangent space to $\Sigma $ at this point, {\bf and} they are
invariant under the action of the group, 
\begin{equation}
\label{gract30}\delta _M\Pi ^A={\cal \not L}_{\widetilde{T_M}}\Pi ^A=0\quad
, 
\end{equation}
where ${\cal \not L}_{\widetilde{T_M}}$ denotes the Lie derivative with
respect to the induced vector field $\widetilde{T_M}$. The WZ-form $\left(
WZ\right) $ on $\Sigma $ can be expanded in this basis,%
$$
\left( WZ\right) =\frac 1{\left( p+1\right) !}\Pi ^{A_1}\cdots \Pi
^{A_{p+1}}\cdot \left( WZ\right) _{A_{p+1}\cdots A_1}\left( \phi \right)
\quad , 
$$
with pull-back 
\begin{equation}
\label{gr333}emb^{*}\left( WZ\right) =\frac 1{\left( p+1\right) !}dx^{\mu
_1}\cdots dx^{\mu _{p+1}}\cdot \Pi _{,\mu _1}^{A_1}\cdots \Pi _{,\mu
_{p+1}}^{A_{p+1}}\cdot \left( WZ\right) _{A_{p+1}\cdots A_1}\left( \phi
\right) \quad ; 
\end{equation}
since $dx^{\mu _1}\cdots dx^{\mu _{p+1}}$ is proportional to the canonical
volume form $\omega _0$ with respect to the coordinates $\left( x^\mu
\right) $ on $W$,%
$$
dx^{\mu _1}\cdots dx^{\mu _{p+1}}=\epsilon ^{\mu _1\cdots \mu _{p+1}}\cdot
\omega _0\quad ,\quad \omega _0=dx^0\cdots dx^p\quad , 
$$
we find that $emb^{*}\left( WZ\right) =\omega _0\cdot {\cal L}_{WZ}$, where 
\begin{equation}
\label{gract319}{\cal L}_{WZ}=\frac 1{\left( p+1\right) !}\,\epsilon ^{\mu
_1\cdots \mu _{p+1}}\cdot \,\Pi _{,\mu _1}^{A_1}\cdots \Pi _{,\mu
_{p+1}}^{A_{p+1}}\cdot \,\left( WZ\right) _{A_{p+1}\cdots A_1}\left( \phi
\right) \quad . 
\end{equation}
Note that we have tacitly used the superspace summation conventions on the
indices $M_i$, which, of course, does not affect the validity of the results
to be shown.

Semi-invariance of the WZ-term then implies that for every generator $T_M$
of $G$ there exists a $p$-form $\Delta _M$ on $\Sigma $ such that 
\begin{equation}
\label{gract31}\delta _M\left( WZ\right) =d\Delta _M\quad . 
\end{equation}
This implies that%
$$
\omega _0\cdot \delta _M{\cal L}_{WZ}=\delta _M\left[ emb^{*}\left(
WZ\right) \right] \;=\;emb^{*}\delta _M\left( WZ\right) =emb^{*}d\Delta _M= 
$$
\begin{equation}
\label{gract32}=d\left( emb^{*}\Delta _M\right) \quad . 
\end{equation}
Expanding $\Delta _M$ in the LI-basis we can compute $d\left( emb^{*}\Delta
_M\right) =$%
$$
=\omega _0\cdot \frac 1{p!}\epsilon ^{\mu _1\cdots \mu _{p+1}}\cdot \partial
_{\mu _1}\,\left[ \Pi _{,\mu _2}^{A_2}\cdots \Pi _{,\mu
_{p+1}}^{A_{p+1}}\cdot \,\Delta _{MA_{p+1}\cdots A_2}\right] \quad , 
$$
and comparison with (\ref{gract32}) then yields%
$$
\delta _M{\cal L}_{WZ}=\partial _\mu U_M^\mu \quad , 
$$
\begin{equation}
\label{gract33}U_M^\mu =\frac 1{p!}\epsilon ^{\mu \mu _2\cdots \mu
_{p+1}}\cdot \,\left[ \Pi _{,\mu _2}^{A_2}\cdots \Pi _{,\mu
_{p+1}}^{A_{p+1}}\cdot \,\Delta _{MA_{p+1}\cdots A_2}\right] \quad . 
\end{equation}
In particular, for $\mu =0$ we obtain%
$$
U_M^0=\frac 1{p!}\epsilon ^{0\mu _2\cdots \mu _{p+1}}\cdot \,\left[ \Pi
_{,\mu _2}^{A_2}\cdots \Pi _{,\mu _{p+1}}^{A_{p+1}}\cdot \,\Delta
_{MA_{p+1}\cdots A_2}\right] \quad , 
$$
from which it is seen that $U_M^0$ {\bf cannot} contain $\Pi _{,0}^A$, due
to the antisymmetry of the $\epsilon $-tensor. Reexpanding the forms $\Pi ^A$
in the coordinate basis $d\phi ^M$ gives%
$$
\Pi ^A=\Pi _N^Ad\phi ^N\quad ,\quad \Pi _{,\mu }^A=\Pi _N^A\phi _{,\mu
}^N\quad , 
$$
from which we see that $U_M^0$ cannot contain $\phi _{,0}^N=\dot \phi ^N$
either. This point is crucial in light of our previous considerations, of
course, since, if we now assume, that the equations%
$$
\Lambda _M=\frac{\partial {\cal L}}{\partial \dot \phi ^M}\quad ,\quad \mbox{%
for\quad }{\cal L}={\cal L}_0+{\cal L}_{WZ} 
$$
are invertible with respect to $\dot \phi ^N$, then $\dot \phi ^M=\Phi
^M\left( \phi ,\partial _r\phi ,\Lambda \right) $ for $r=1,\ldots ,p$, and
after performing the Legendre transformation $\left( \phi ^M,\dot \phi
^N\right) \rightarrow \left( \phi ^M,\Lambda _N\right) $ we have%
$$
U_M^s=U_M^s\left( \phi ,\Phi ^M\left( \phi ,\partial _r\phi ,\Lambda \right)
,\partial _t\phi \right) =\widehat{U_M^s}\left( \phi ,\partial _t\phi
,\Lambda ,\right) \quad ;\quad s=1,\ldots ,p\,;\;r,t\neq s\quad , 
$$
but 
\begin{equation}
\label{gr001}U_M^0=\widehat{U_M^0}\left( \phi ,\partial _r\phi \right) \quad
;\quad r=1,\ldots ,p\quad . 
\end{equation}
Therefore we now have Poisson brackets 
\begin{equation}
\label{gract34}\left\{ j_M^0,\widehat{U_N^0}\right\} =\delta _M\phi ^K\cdot
\left\{ \Lambda _K,\widehat{U_N^0}\right\} \quad , 
\end{equation}
\begin{equation}
\label{gract35}\left\{ \widehat{U_M^0},\widehat{U_N^0}\right\} =0\quad . 
\end{equation}
This point being clarified we omit the ''hats'' on $\widehat{U_N^0}$ from
now on, it being understood that it is the ''hatted'' version that appears
in a Poisson bracket.

Referring to (\ref{gract29}) we can now state that the algebra of the
charges $Q_M$ closes to a linear combination of the $Q_M$ and the new
charges $Z_{MN}$, 
\begin{equation}
\label{cl1}\left\{ Q_M,Q_N\right\} _{PB}=\pm C_{MN}^{\;K}\cdot
Q_K\;+\;Z_{MN}\quad .
\end{equation}
Furthermore, due to (\ref{gract35}), we have 
\begin{equation}
\label{cll}\left\{ S_{MN}^0,S_{M^{\prime }N^{\prime }}^0\right\} =0\quad ,
\end{equation}
and therefore 
\begin{equation}
\label{cl2}\left\{ Z_{MN},Z_{M^{\prime }N^{\prime }}\right\} =0\quad ,
\end{equation}
i.e. the mutual algebra of the new charges $Z_{MN}$ also closes into the
extension generated by $\left( Q_M,Z_{MN}\right) $. But what about the
algebra of $\left\{ Q_K,Z_{MN}\right\} _{PB}$ ? We now examine under which
conditions this expression yields a linear combination of $\left\{
Q_M,Z_{MN}\right\} _{PB}$.

\subsubsection{Topological currents\quad \quad \label{TopCurrents}}

Consider the object $S_{MN}^\mu $ defined in (\ref{gract24}), 
\begin{equation}
\label{sch}S_{MN}^\mu \left( t,\sigma \right) =\delta _MU_N^\mu -\delta
_NU_M^\mu \pm C_{MN}^{\;K}\cdot U_K^\mu \quad . 
\end{equation}
Using the form of $U_M^\mu $ given in (\ref{gract33}) and taking into
account that $\delta _M\Pi _{,\nu }^N=0$ we have%
$$
S_{MN}^\mu =\frac 1{p!}\epsilon ^{\mu \mu _1\cdots \mu _p}\,\,\Pi _{,\mu
_1}^{A_1}\cdots \Pi _{,\mu _p}^{A_p}\;\times 
$$
\begin{equation}
\label{gract36}\times \;\left[ \,\delta _M\Delta _{NA_p\cdots A_1}-\delta
_N\Delta _{MA_p\cdots A_1}\pm C_{MN}^{\;K}\cdot \Delta _{KA_p\cdots
A_1}\right] \quad . 
\end{equation}
For the sake of simplicity we define the expression 
\begin{equation}
\label{gract37}\tilde R_{MNA_p\cdots A_1}:=\left[ \,\delta _M\Delta
_N-\delta _N\Delta _M\pm C_{MN}^{\;K}\cdot \Delta _K\right] _{A_p\cdots
A_1}\quad , 
\end{equation}
so that%
$$
S_{MN}^\mu =\frac 1{p!}\epsilon ^{\mu \mu _1\cdots \mu _p}\,\,\Pi _{,\mu
_1}^{A_1}\cdots \Pi _{,\mu _p}^{A_p}\cdot \tilde R_{MNA_p\cdots A_1}\quad , 
$$
and rewrite this form in the coordinate basis $\left( d\phi ^N\right) $, $%
\Pi ^A=\Pi _N^Ad\phi ^N$, which yields 
\begin{equation}
\label{gract38}S_{MN}^\mu =\frac 1{p!}\epsilon ^{\mu \mu _1\cdots \mu
_p}\,\,\phi _{,\mu _1}^{N_1}\cdots \phi _{,\mu _p}^{N_p}\cdot R_{MNN_p\cdots
N_1}\quad , 
\end{equation}
with the new components 
\begin{equation}
\label{gract39}R_{MNN_p\cdots N_1}=\Pi _{N_1}^{A_1}\cdots \Pi
_{N_p}^{A_p}\cdot \tilde R_{MNA_p\cdots A_1}\quad . 
\end{equation}
We note that $R_{MNN_p\cdots N_1}$ is a function of the fields $\phi $ only.
Appealing to (\ref{gract38}) we now define the identically conserved \cite
{Azca} {\it topological currents}

\begin{equation}
\label{gract40}j_T^{\mu M_1\cdots M_p}:=\epsilon ^{\mu \mu _1\cdots \mu
_p}\,\,\phi _{,\mu _1}^{M_1}\cdots \phi _{,\mu _p}^{M_p}\quad , 
\end{equation}
and the {\it topological charges} 
\begin{equation}
\label{gract401}T^{M_1\cdots M_p}:=\int\limits_{W\left( t\right) }d^p\sigma
\cdot j_T^{0M_1\cdots M_p}\quad , 
\end{equation}
which are conserved due to $\partial _\mu j_T^{\mu M_1\cdots M_p}=0$. The
topological charges $T^{M_1\cdots M_p}$ are invariant under the group
action, 
\begin{equation}
\label{da}\delta _KT^{M_1\cdots M_p}=0\quad , 
\end{equation}
see (\ref{ko5}) below. We write (\ref{gract38}) as 
\begin{equation}
\label{gract41}S_{MN}^\mu =\frac 1{p!}j_T^{\mu N_1\cdots N_p}\cdot
R_{MNN_p\cdots N_1}=:\,j_T^\mu \bullet R_{MN}\quad , 
\end{equation}
then the charges $Z_{MN}$ take the form 
\begin{equation}
\label{gr401}Z_{MN}=\int\limits_{W\left( t\right) }d^p\sigma \cdot
S_{MN}^0=\int\limits_{W\left( t\right) }d^p\sigma \;j_T^0\bullet R_{MN}\quad
; 
\end{equation}
for {\bf constant} $R_{MNN_p\cdots N_1}$ this is 
\begin{equation}
\label{gract262}Z_{MN}=T\bullet R_{MN}\quad . 
\end{equation}

Now we can turn to the bracket $\left\{ Q_K,Z_{MN}\right\} $; a computation
yields 
\begin{equation}
\label{gract49}\left\{ Q_K,Z_{MN}\right\} =-\,\int\limits_{W\left( t\right)
}d^p\sigma \cdot \delta _K\left[ j_T^0\bullet R_{MN}\right] \quad . 
\end{equation}
It is clear that this can never close into an expression involving the
charges $Q_M$, since this would require the occurence of $j_M^0=\delta
_M\phi ^N\Lambda _N$ in the integrand, but the integrand contains no
canonical momenta (recall that $j_T^0$ contains no time derivatives of
fields, and $R_{MN}$ contains no field derivatives at all). Hence, at best
the left hand side can close into a linear combination of the new charges $%
Z_{M^{\prime }N^{\prime }}$. If we now look at (\ref{gr401}) we see that
requiring that (\ref{gract49}) be a linear combination of $Z_{M^{\prime
}N^{\prime }}$ is equivalent to demanding that 
\begin{equation}
\label{gract505}\delta _K\left[ j_T^0\bullet R_{MN}\right] =-\frac
12B_{KMN}^{M^{\prime }N^{\prime }}\cdot j_T^0\bullet R_{M^{\prime }N^{\prime
}}\;+\;\cdots \quad , 
\end{equation}
where $\cdots $ denote possible surface terms, and where $B_{KMN}^{M^{\prime
}N^{\prime }}$ are {\bf constant}; the factor $\frac 12$ is due to the
antisymmetry of $R_{MN}$ in $M$ and $N$. (\ref{gract49}) then reads 
\begin{equation}
\label{gract501}\left\{ Q_K,Z_{MN}\right\} =\frac 12B_{KMN}^{M^{\prime
}N^{\prime }}\cdot Z_{M^{\prime }N^{\prime }}\quad . 
\end{equation}
(\ref{cl1}, \ref{cl2}) and (\ref{gract49} - \ref{gract501}) now tell us that
the algebra of the conserved charges $Q_K,Z_{MN}$ closes if and only if (\ref
{gract505}) holds.

\subsubsection{When are the charges $Z_{MN}$ central ?}

This can be read off from (\ref{gract501}): The charges $Z_{MN}$ are {\it %
central}, i.e. they commute with all other elements in the algebra, {\bf iff}
all coefficients $B_{KMN}^{M^{\prime }N^{\prime }}$ vanish; according to (%
\ref{gract505}) this is true {\bf iff} 
\begin{equation}
\label{ko2}\delta _K\left[ j_T^0\bullet R_{MN}\right] =\left( \mbox{globally
defined smooth surface term}\right) \quad . 
\end{equation}
Let us now examine 
\begin{equation}
\label{ko3}\delta _Kj_T^{0M_1\cdots M_p}=\sum_{k=1}^p\partial _{\mu
_k}\left[ \frac 1{p!}\epsilon ^{0\mu _1\cdots \mu _k\cdots \mu _p}\cdot \phi
_{,\mu _1}^{M_1}\cdots \delta _K\phi ^{M_k}\cdots \phi _{,\mu
_p}^{M_p}\right] \quad . 
\end{equation}
We take the point of view that the expression in square brackets is smooth
and globally defined (since $\delta _K\phi ^{M_k}$ amounts to a derivative
of the field $\phi ^{M_k}$ which can be smoothy continued over the whole of $%
W$) so that its integral over $W\left( t\right) $ indeed vanishes, on using
Stokes' theorem. This means that $\delta _K\left[ j_T^0\bullet R_{MN}\right] 
$ is a surface term, {\bf provided} that $R_{MN}$ are constant. A {\bf %
sufficient} condition for the charges $Z_{MN}$ to be {\bf central} is
therefore that 
\begin{equation}
\label{ko4}R_{MNN_p\cdots N_1}\left( \phi \right) =const.=R_{MNN_p\cdots
N_1}\quad , 
\end{equation}
where $R_{MNN_p\cdots N_1}$ are the components of $R_{MN}$ in the coordinate
basis $\left( d\phi ^N\right) $.

As an aside we remark that (\ref{ko3}) implies that the topological charges $%
T^{M_1\cdots M_p}$ are invariant under the group action, 
\begin{equation}
\label{ko5}\delta _KT^{M_1\cdots M_p}=0\quad . 
\end{equation}

We now have (see (\ref{gract262})) $Z_{MN}=T\bullet R_{MN}$, and the
non-vanishing brackets of our extended algebra then read%
$$
\left\{ Q_M,Q_N\right\} =C_{MN}^{\;K}\cdot Q_K\;+\;T\bullet R_{MN}\quad , 
$$
\begin{equation}
\label{gract53}\left\{ Q_K,T\bullet R_{MN}\right\} =\frac
12B_{KMN}^{M^{\prime }N^{\prime }}\cdot T\bullet R_{M^{\prime }N^{\prime
}}\quad , 
\end{equation}
and the charges $T\bullet R_{MN}$ are all {\bf central}.

\subsection{Summary}

At this point it is appropriate to summarize the results we have obtained so
far in the form of three theorems.

\subsubsection{Theorem 1}

The Noether currents satisfy the Poisson bracket algebra, possibly up to a
sign, 
\begin{equation}
\label{th1}\left\{ j_M^0\left( t,\sigma \right) ,j_N^0\left( t,\sigma
^{\prime }\right) \right\} _{PB}\;=\;\pm C_{MN}^{\;K}\cdot j_K^0\left(
t,\sigma \right) \cdot \delta \left( \sigma -\sigma ^{\prime }\right) \quad ,
\end{equation}
{\bf regardless} of whether the Lagrangian ${\cal L}$ is {\bf invariant or
not}. $\pm $ refers to a left/right action. The once integrated version is 
\begin{equation}
\label{th2}\left\{ Q_M,j_N^0\right\} _{PB}=\;\pm j_K^0\cdot ad\left(
T_M\right) _{\;N}^K\quad ,
\end{equation}
where $ad\left( T\right) $ denotes the adjoint representation of the Lie
algebra element $T$. This implies that the total Noether currents span the
adjoint representation of $G$ in the case of a left action.

Double integration of the current algebra yields the algebra of the
generators of $G$, possibly up to a sign, 
\begin{equation}
\label{th3}\left\{ Q_M,Q_N\right\} _{PB}=\pm C_{MN}^{\;K}\cdot Q_K\quad .
\end{equation}

\subsubsection{Theorem 2}

Assume that the Lagrangian ${\cal L}$ is {\bf semi-invariant} under the
action of $G$, i.e. $\delta _M{\cal L}=\partial _\mu U_M^\mu $ for functions 
$U_M^\mu =U_M^\mu \left( \phi ,\partial _\nu \phi \right) $ of the fields
and its derivatives {\bf on-shell and off-shell}; that the action of $G$ on
canonical momenta $\Lambda _i={\cal L}_{\dot \phi ^i}$ is defined by (\ref
{gract17}); and that surface integrals with smooth integrands may be
neglected. Then

\begin{enumerate}
\item  The modified currents 
\begin{equation}
\label{th5}\widetilde{j_M^\mu }=j_M^\mu -U_M^\mu \quad ,
\end{equation}
where $j_M^\mu \,$ are the Noether currents associated with ${\cal L}$, are
conserved, 
\begin{equation}
\label{th6}\partial _\mu \widetilde{j_M^\mu }=0\quad .
\end{equation}

\item  Double integration of the Poisson bracket algebra yields 
\begin{equation}
\label{th7}\left\{ Q_M,Q_N\right\} _{PB}=\pm C_{MN}^{\;K}\cdot
Q_K\;+\;Z_{MN}\;+\int\limits_{W\left( t\right) }d^p\sigma \,d^p\sigma
^{\prime }\cdot \left\{ \widehat{U_M^0}\left( t,\sigma \right) ,\widehat{%
U_N^0}\left( t,\sigma ^{\prime }\right) \right\} \quad ,
\end{equation}
where 
\begin{equation}
\label{th8}Z_{MN}\left( t\right) :=\int\limits_{W\left( t\right) }d^p\sigma
\cdot S_{MN}^0\left( t,\sigma \right) \quad ,
\end{equation}
and 
\begin{equation}
\label{th9}S_{MN}^\mu \left( t,\sigma \right) =\delta _MU_N^\mu -\delta
_NU_M^\mu \pm C_{MN}^{\;K}\cdot U_K^\mu \quad .
\end{equation}
Due to 
\begin{equation}
\label{th10}\partial _\mu S_{MN}^\mu =0
\end{equation}
the ''charges'' $Z_{MN}$ are {\bf conserved}.
\end{enumerate}

\subsubsection{Theorem 3}

Let those directions of the hypersurfaces $W\left( t\right) $ which are not
infinitely extended be closed. Let the target space $\Sigma $ be the group $%
G $ itself; let the semi-invariant piece ${\cal L}_1={\cal L}_{WZ}$ in the
Lagrangian be the pull-back of a target space $\left( p+1\right) $-form to
the worldvolume $W$, which transforms under $G$ according to $\delta _M{\cal %
L}_{WZ}=\partial _\mu U_M^\mu $, with 
\begin{equation}
\label{th11}U_M^\mu =\frac 1{p!}\epsilon ^{\mu \mu _2\cdots \mu _{p+1}}\cdot
\,\left[ \Pi _{,\mu _2}^{M_2}\cdots \Pi _{,\mu _{p+1}}^{M_{p+1}}\cdot
\,\Delta _{MM_{p+1}\cdots M_2}\right] \quad , 
\end{equation}
where $\Delta _{MM_{p+1}\cdots M_2}$ are the components of $\dim G$ $p$%
-forms $\Delta _M$ in a left-invariant basis $\left( \Pi ^M\right) $. Let
the action of $G$ on canonical momenta $\Lambda _i={\cal L}_{\dot \phi ^i}$
be defined according to (\ref{gract17}). Then

\begin{enumerate}
\item  The Poisson bracket algebra of the Noether charges $Q_M$ and the
charges $Z_{MN}$ closes {\bf iff} 
\begin{equation}
\label{th12}\delta _K\left[ j_T^0\bullet R_{MN}\right] =-\frac
12B_{KMN}^{M^{\prime }N^{\prime }}\cdot j_T^0\bullet R_{M^{\prime }N^{\prime
}}\;+\;\cdots \quad ,
\end{equation}
where $\cdots $ denote possible surface terms, $B_{KMN}^{M^{\prime
}N^{\prime }}$ are {\bf constant}, and where 
\begin{equation}
\label{th13}R_{MNN_p\cdots N_1}=\Pi _{N_1}^{A_1}\cdots \Pi _{N_p}^{A_p}\cdot
\left[ \,\delta _M\Delta _N-\delta _N\Delta _M\pm C_{MN}^{\;K}\cdot \Delta
_K\right] _{A_p\cdots A_1}\quad .
\end{equation}
The extended algebra then reads 
\begin{equation}
\label{th14}\left\{ Q_M,Q_N\right\} _{PB}=\pm C_{MN}^{\;K}\cdot
Q_K\;+\;Z_{MN}\quad ,
\end{equation}
\begin{equation}
\label{th15}\left\{ Q_K,Z_{MN}\right\} =\frac 12B_{KMN}^{M^{\prime
}N^{\prime }}\cdot Z_{M^{\prime }N^{\prime }}\quad ,
\end{equation}
\begin{equation}
\label{th16}\left\{ Z_{MN},Z_{M^{\prime }N^{\prime }}\right\} _{PB}=0\quad .
\end{equation}

\item  A {\bf sufficient} condition for the charges $Z_{MN}$ to be {\bf %
central} is that 
\begin{equation}
\label{th17}R_{MNN_p\cdots N_1}\left( \phi \right) =const.=R_{MNN_p\cdots
N_1}\quad .
\end{equation}

\item  The topological charges $T^{M_1\cdots M_p}$ are invariant under the
group action, 
\begin{equation}
\label{th19}\delta _KT^{M_1\cdots M_p}=0\quad .
\end{equation}
\end{enumerate}

\subsubsection{Corollary}

If $R_{MNN_p\cdots N_1}=const.$, then all charges $Z_{MN}$ are central, and
are linear combinations of the topological charges $T^{M_1\cdots M_p}$, 
\begin{equation}
\label{th18}Z_{MN}=T\bullet R_{MN}\quad . 
\end{equation}

\subsection{Lagrangians including (Abelian) Gauge fields \label{Eich}}

\subsubsection{Structure of the Lagrangian}

Now let us study the case when the Lagrangian ${\cal L}$ contains additional
degrees of freedom in the form of an Abelian $\left( q-1\right) $-form gauge
potential $A_{\mu _1\ldots \mu _{q-1}}$, $q\le p$, that is defined on the 
{\bf worldvolume}. A priori, the group $G$ acts on the target space $\Sigma $
and there is no reason why $A$ should be involved in the transformation of
fields on $\Sigma $, but that is what we now impose on $A$, since it is the
situation that occurs when the Lagrangian describes $D$-$p$-branes, which we
want to study later. To this end, we assume that on the target space there
exists a $q$-form potential $B=\frac 1{q!}\Pi ^{C_q}\cdots \Pi
^{C_1}B_{C_1\cdots C_q}$, with an associated $\left( q+1\right) $-form field
strength $H=dB$. The field strength $H$ is taken to be invariant under the
action of $G$, i.e. $\delta _MH=0$. This implies that locally 
\begin{equation}
\label{ga0}\delta _MB=d\Delta _M\quad , 
\end{equation}
with $\dim G$ $\left( q-1\right) $-forms%
$$
\Delta _M=\frac 1{\left( q-1\right) !}\Pi ^{A_q}\cdots \Pi ^{A_2}\widetilde{%
\Delta _{MA_2\cdots A_q}}=\frac 1{\left( q-1\right) !}d\phi ^{A_q}\cdots
d\phi ^{A_2}\,\Delta _{MA_2\cdots A_q}\quad , 
$$
where we have used a tilde to distinguish the components of $\Delta _M$ with
respect to the LI-basis $\left( \Pi ^A\right) $ from the components in the
coordinate basis $\left( d\phi ^M\right) $, which we shall need later. $%
A_{\mu _1\ldots \mu _{q-1}}$ are therefore ${{p+1} \choose {q-1}}$ additional
degrees of freedom involved in the dynamics; it is assumed, however, that $%
A_{\mu _1\ldots \mu _{q-1}}$ enters the Lagrangian {\bf only} via the field
strengths $F_{\mu _1\ldots \mu _q}=q\cdot \partial _{[\mu _1}A_{\mu _2\ldots
\mu _q]}$. The Lagrangian again splits into an invariant piece ${\cal L}_0$
and a semi-invariant piece ${\cal L}_{WZ}$, where ${\cal L}_0$ takes the
form 
\begin{equation}
\label{ga2}{\cal L}_0={\cal L}_0\left( \phi ,\partial _\mu \phi ,\widehat{%
F_{\mu _1\ldots \mu _q}}\right) \quad ,\quad \widehat{F_{\mu _1\ldots \mu _q}%
}=F_{\mu _1\ldots \mu _q}-\left( emb^{*}B\right) _{\mu _1\ldots \mu _q}\quad
; 
\end{equation}
in order to have ${\cal L}_0$ invariant we {\bf impose} the transformation
behaviour 
\begin{equation}
\label{ga3}\delta _MA=emb^{*}\Delta _M\quad ,\mbox{\quad }\left( \delta
_MA\right) _{\mu _2\ldots \mu _q}=\phi _{,\mu _2}^{A_q}\cdots \phi _{,\mu
_q}^{A_2}\,\Delta _{MA_2\cdots A_q}\left( \phi \right) 
\end{equation}
on $A$. Since 
\begin{equation}
\label{ga4}\delta _M\widehat{F}=\delta _M\left[ dA-emb^{*}B\right] =\left[
d\delta _MA-emb^{*}d\Delta _M\right] =0\quad , 
\end{equation}
this is sufficient to have an invariant ${\cal L}_0$. As for the
semi-invariant part ${\cal L}_{WZ}$, we assume the following: 
\begin{equation}
\label{ga5}{\cal L}_{WZ}={\cal L}_{WZ}\left( \phi ,\partial _\mu \phi ,%
\widehat{F_{\mu _1\ldots \mu _q}}\right) \quad , 
\end{equation}
with transformation behaviour $\delta _M{\cal L}_{WZ}=\partial _\mu U_M^\mu $%
, where $U_M^\mu =U_M^\mu \left( \phi ,\partial _\mu \phi ,F_{\mu \nu
}\right) $, {\bf but} 
\begin{equation}
\label{ga6}\frac{\partial U_M^\mu }{\partial \partial _\nu \phi }=0\quad
\quad \mbox{if\quad }\mu =\nu \quad ;\quad \frac{\partial U_M^\mu }{\partial
F_{\nu _1\ldots \nu _q}}=0\quad \quad \mbox{if\quad }\mu \in \left\{ \nu
_1,\ldots ,\nu _q\right\} \quad . 
\end{equation}
Note the absence of a hat in the field $F$ in the definition of the field
content of $U_M^\mu $.

Now we define canonical momenta 
\begin{equation}
\label{ga61}\Lambda _N=\frac{\partial {\cal L}}{\partial \partial _0\phi ^N}%
\quad ,\quad \Lambda ^{\nu _2\ldots \nu _q}=\frac{\partial {\cal L}}{%
\partial \partial _0A_{\nu _2\ldots \nu _q}}\quad . 
\end{equation}

\subsubsection{Constraints on the gauge field degrees of freedom}

The fact that we are dealing with a gauge field $A_{\mu _1\ldots \mu _{q-1}}$
as dynamical degrees of freedom makes itself manifest in the form of {\bf %
constraints} that are imposed on the dynamics \cite{Govaerts}: Using the
formula 
\begin{equation}
\label{ga7}\frac{\partial {\cal L}}{\partial \partial _{\nu _1}A_{\nu
_2\ldots \nu _q}}=\frac 1{\left( q-1\right) !}\frac{\partial {\cal L}}{%
\partial F_{\nu _1\ldots \nu _q}} 
\end{equation}
we see that, due to the antisymmetry of $F$, ${\cal L}$ cannot contain $%
\partial _0A_{\nu _2\ldots \nu _q}$, whenever one of the $\nu _2,\ldots ,\nu
_q$ is zero; this implies that the canonical momenta 
\begin{equation}
\label{ga8}\Lambda ^{\nu _2\ldots \nu _q}=0\quad \quad \mbox{for\quad }0\in
\left\{ \nu _2,\ldots \nu _q\right\} \quad , 
\end{equation}
i.e. they vanish identically. The number of independent constraints (\ref
{ga8}) is ${p \choose {q-2}}$. The second set of constraints follows from the
equations of motion for $A_{\nu _2\ldots \nu _q}$: They are given by%
$$
\left( eq\right) ^{\nu _2\ldots \nu _q}:=\frac{\partial {\cal L}}{\partial
A_{\nu _2\ldots \nu _q}}-\partial _0\Lambda ^{\nu _2\ldots \nu _q}-\partial
_r\frac{\partial {\cal L}}{\partial \partial _rA_{\nu _2\ldots \nu _q}}%
=0\quad , 
$$
where the sum over $r$ ranges from $1$ to $p$. The first term on the RHS
vanishes since ${\cal L}$ contains no $A_{\nu _2\ldots \nu _q}$; the second
one vanishes if we choose one of the $\nu $'s to be equal to zero, say $\nu
_2$. On using (\ref{ga7}) we have $\frac{\partial {\cal L}}{\partial
\partial _rA_{0\nu _3\ldots \nu _q}}=-\frac{\partial {\cal L}}{\partial
\partial _0A_{r\nu _3\ldots \nu _q}}$, so we get 
\begin{equation}
\label{ga9}\partial _r\Lambda ^{r\nu _3\ldots \nu _q}=0\quad ,\quad \nu
_3,\ldots ,\nu _q\mbox{\quad arbitrary.} 
\end{equation}
This yields a number of another ${p \choose {q-2}}$ constraints.

These constraints are not on an equal footing, however; as can be seen from
the above arguments, the first set (\ref{ga8}) holds before any equations of
motion are considered, and therefore amounts to a reduction of phase space
to a submanifold of the original phase space of codimension ${p \choose {q-2}}$;
in Dirac's terminology this is a set of {\it primary constraints}. The
second set (\ref{ga9}) comes into play only {\bf on-shell}, i.e. on using
equations of motion, and is called a set of {\it secondary constraints}. We
shall not use Dirac's machinery for handling these constraints here, but
shall work with Poisson brackets instead; in this case, however, it is
crucial to impose (\ref{ga8}, \ref{ga9}) {\bf not before} all Poisson
brackets have been worked out, otherwise we would obtain wrong results.

After these remarks let us now study the Poisson bracket algebra of the
Noether currents. The Noether currents are 
\begin{equation}
\label{ga10}j_M^\mu =\delta _M\phi ^K\cdot \frac{\partial {\cal L}}{\partial
\partial _\mu \phi ^K}+\frac 1{\left( q-1\right) !}\delta _MA_{\nu _2\ldots
\nu _q}\cdot \frac{\partial {\cal L}}{\partial \partial _\mu A_{\nu _2\ldots
\nu _q}}\quad , 
\end{equation}
\begin{equation}
\label{ga11}j_M^0=\delta _M\phi ^K\cdot \Lambda _K+\frac 1{\left( q-1\right)
!}\delta _MA_{\nu _2\ldots \nu _q}\cdot \Lambda ^{\nu _2\ldots \nu _q}\quad
. 
\end{equation}

\subsubsection{Algebra of Noether currents}

In working out brackets $\left\{ j_M^0,j_N^0\right\} $ we make use of the
fact that $\delta _M\phi ^K$ is a function of the fields $\phi $ only,
therefore the brackets $\left\{ \delta _M\phi ^K,\Lambda ^{\nu _2\ldots \nu
_q}\right\} $ vanish; and that $\delta _MA_{\nu _2\ldots \nu _q}$ is a
function of the fields $\phi $ and their derivatives $\partial _\mu \phi $
only, see (\ref{ga3}), therefore the brackets $\left\{ \delta _MA_\nu
,\Lambda ^{\nu _2\ldots \nu _q}\right\} $ vanish. The computation then yields%
$$
\left\{ j_M^0\left( t,\sigma \right) ,j_N^0\left( t,\sigma ^{\prime }\right)
\right\} =\pm C_{MN}^{\;K}\cdot j_K^0\cdot \delta \left( \sigma -\sigma
^{\prime }\right) \;+\;\frac 1{\left( q-1\right) !}\left[ \,\mp
C_{MN}^{\;K}\,\delta _KA_{\nu _2\ldots \nu _q}\cdot \delta \left( \sigma
-\sigma ^{\prime }\right) \;+\right. 
$$
\begin{equation}
\label{ga12}\left. +\;\delta _M\phi ^K\cdot \left\{ \Lambda _K,\delta
_NA_{\nu _2\ldots \nu _q}\right\} \;-\;\delta _N\phi ^K\cdot \left\{ \Lambda
_K,\delta _MA_{\nu _2\ldots \nu _q}\right\} \right] \cdot \Lambda ^{\nu
_2\ldots \nu _q}\quad . 
\end{equation}
This can be written as%
$$
\left\{ j_M^0,j_N^0\right\} =\pm C_{MN}^{\;K}\cdot j_K^0\cdot \delta \left(
\sigma -\sigma ^{\prime }\right) \;+\; 
$$
\begin{equation}
\label{ga13}+\;\frac 1{\left( q-1\right) !}\left[ \,\left( -\delta _M\delta
_N+\delta _N\delta _M\mp C_{MN}^{\;K}\,\cdot \delta _K\right) A_{\nu
_2\ldots \nu _q}\right] \Lambda ^{\nu _2\ldots \nu _q}\cdot \delta \left(
\sigma -\sigma ^{\prime }\right) \quad \cdots \quad , 
\end{equation}
where $\cdots $ denotes surface terms.

The first term is just what we have expected; $\pm $ again refers to a
left/right action. Since $\delta _MA=emb^{*}\Delta _M$ we have $\left(
-\delta _M\delta _N+\delta _N\delta _M\mp C_{MN}^{\;K}\,\cdot \delta
_K\right) A=$%
$$
=\;emb^{*}\left( -\delta _M\Delta _N+\delta _N\Delta _M\mp
C_{MN}^{\;K}\,\Delta _K\right) \quad , 
$$
where the expression in the brackets 
\begin{equation}
\label{ga14}-\delta _M\Delta _N+\delta _N\Delta _M\mp C_{MN}^{\;K}\,\Delta
_K\;=:\;S\left( \Delta \right) _{MN}\; 
\end{equation}
measures the {\bf deviation} of the forms $\Delta _M$ from transforming as a
multiplet under the adjoint representation of the group $G$; this is seen
from 
\begin{equation}
\label{ga15}T_M\cdot \Delta _N=\mp \,\Delta _K\cdot ad\left( T_M\right)
_{\;N}^K\;-\;S\left( \Delta \right) _{MN}\quad , 
\end{equation}
where the point denotes the action of the ''abstract'' generator $T_M$ on
the component $\Delta _N$ according to $T_M\cdot \Delta _N=\left[ \delta
_M,\Delta _N\right] $.

We now introduce the notation 
\begin{equation}
\label{ga16}\left[ emb^{*}S\left( \Delta \right) _{MN}\right] _{\nu _2\ldots
\nu _q}=:S\left( \Delta \right) _{MN\nu _2\ldots \nu _q}\quad ,
\end{equation}
\begin{equation}
\label{ga17}\frac 1{\left( q-1\right) !}S\left( \Delta \right) _{MN\nu
_2\ldots \nu _q}\Lambda ^{\nu _2\ldots \nu _q}=:S\left( \Delta \right)
_{MN}\bullet \Lambda ^{gauge}\quad ,
\end{equation}
then (\ref{ga13}) reads 
\begin{equation}
\label{ga18}\left\{ j_M^0,j_N^0\right\} =\left[ \pm C_{MN}^{\;K}\cdot
j_K^0\;+\;S\left( \Delta \right) _{MN}\bullet \Lambda ^{gauge}\right] \cdot
\delta \left( \sigma -\sigma ^{\prime }\right) \quad .
\end{equation}
Let us define 
\begin{equation}
\label{ga19}Q_M=\int\limits_{W\left( t\right) }d^p\sigma \cdot j_M^0\quad
,\quad Y_{MN}\left( t\right) =\int\limits_{W\left( t\right) }d^p\sigma \cdot
S\left( \Delta \right) _{MN}\bullet \Lambda ^{gauge}\quad ,
\end{equation}
then the once integrated version of (\ref{ga18}) is 
\begin{equation}
\label{ga20}\left\{ Q_M,j_N^0\right\} =\pm j_K^0\cdot ad\left( T_M\right)
_{\;N}^K\;+\;S\left( \Delta \right) _{MN}\bullet \Lambda ^{gauge}\quad ,
\end{equation}
which defines the action of the generator $T_M$ on the Noether current $j_N^0
$. We see that due to the presence of the $S$-term on the right hand side
the Noether currents now {\bf fail} to transform as a multiplet in the
adjoint representation, as was the case previously.

The twice integrated version is 
\begin{equation}
\label{ga21}\left\{ Q_M,Q_N\right\} =\pm C_{MN}^{\;K}\cdot
Q_K\;+\;Y_{MN}\quad .
\end{equation}
$Q_M$ are conserved when the Lagrangian is invariant under $G$; we need to
check when $Y_{MN}\left( t\right) $ are conserved. To this end we perform $%
\frac d{dt}$ on $Y_{MN}$ in (\ref{ga19}) and assume, for the sake of
convenience, that the hypersurfaces $W\left( t\right) $ do not change shape
as $t$ varies; then the only contribution to $\frac{dY_{MN}}{dt}$ comes from 
$\frac d{dt}\left[ S\left( \Delta \right) _{MN}\bullet \Lambda
^{gauge}\right] $. A calculation then shows that a {\bf sufficient}
condition for the charge $Y_{MN}\left( t\right) $ to be conserved is 
\begin{equation}
\label{ga23}S\left( \Delta \right) _{MNN_q\ldots N_2}\left( \phi \right)
=const.=S\left( \Delta \right) _{MNN_q\ldots N_2}\quad .
\end{equation}
Under the same condition the charges $Y_{MN}$ are seen to be central.

The complete algebra is then 
\begin{equation}
\label{ga25}\left\{ Q_M,Q_N\right\} =\pm C_{MN}^{\;K}\cdot
Q_K\;+\;Y_{MN}\quad ,\quad \left\{ Q_K,Y_{MN}\right\} =\left\{
Y_{MN},Y_{M^{\prime }N^{\prime }}\right\} =0\quad .
\end{equation}

\subsubsection{Algebra of modified currents}

At last then let us determine the general structure of the extended algebra
of the charges associated with the {\bf modified} currents $\widetilde{%
j_M^\mu }=j_M^\mu -U_M^\mu $, given that the WZ-term ${\cal L}_{WZ}$ behaves
as in (\ref{ga5}, \ref{ga6}). We again find that%
$$
\left\{ \widetilde{j_M^0},\widetilde{j_N^0}\right\} =\left\{
j_M^0,j_N^0\right\} -\left\{ U_M^0,j_N^0\right\} -\left\{
j_M^0,U_N^0\right\} \quad , 
$$
with $\left\{ j_M^0,j_N^0\right\} $ given in (\ref{ga18}). $\left\{
j_M^0,U_N^0\right\} $ can be determined using the properties of $U_N^0\,$
given in (\ref{ga6}). Up to surface terms we then find%
$$
\left\{ \widetilde{j_M^0},\widetilde{j_N^0}\right\} \approx \pm
C_{MN}^{\;K}\cdot \widetilde{j_K^0}\cdot \delta \left( \sigma -\sigma
^{\prime }\right) \;+\;\left[ \pm C_{MN}^{\;K}\cdot U_K^0\;+\;S\left( \Delta
\right) _{MN}\bullet \Lambda ^{gauge}\;-\right.  
$$
\begin{equation}
\label{ga26}\left. -\delta _N\phi ^K\frac{\partial U_M^0}{\partial \phi ^K}%
\;+\;\delta _M\phi ^K\frac{\partial U_N^0}{\partial \phi ^K}\right] \cdot
\delta \left( \sigma -\sigma ^{\prime }\right) \quad .
\end{equation}
Analogous to (\ref{sch}) we define 
\begin{equation}
\label{ga27}S\left( U\right) _{MN}=\delta _M\phi ^K\cdot \frac{\partial U_N^0%
}{\partial \phi ^K}-\delta _N\phi ^K\cdot \frac{\partial U_M^0}{\partial
\phi ^K}\pm C_{MN}^{\;K}\cdot U_K^0
\end{equation}
and its integral 
\begin{equation}
\label{ga28}Z_{MN}=\int\limits_{W\left( t\right) }d^p\sigma \cdot S\left(
U\right) _{MN}\quad ,
\end{equation}
and $Y_{MN}\,$ as the integral of $S\left( \Delta \right) _{MN}\bullet
\Lambda ^{gauge}$ over $W\left( t\right) $, according to (\ref{ga19}). Then
double integration of (\ref{ga26}) yields 
\begin{equation}
\label{ga29}\left\{ Q_M,Q_N\right\} =\pm C_{MN}^{\;K}\cdot
Q_K\;+\;Y_{MN}\;+\;Z_{MN}\quad .
\end{equation}
In the case of constant $S\left( \Delta \right) _{MNN_q\ldots N_2}$ (see (%
\ref{ga23})) we can write%
$$
Y_{MN}=\frac 1{\left( q-1\right) !}S\left( \Delta \right) _{MNN_q\ldots
N_2}\int\limits_{W\left( t\right) }d^p\sigma \cdot \phi _{,r_2}^{N_2}\cdots
\phi _{,r_q}^{N_q}\cdot \Lambda ^{r_2\ldots r_q}\,\quad ; 
$$
on the right hand side now there appear charges 
\begin{equation}
\label{ga30}Y_{MN}^{N_2\ldots N_q}:=\int\limits_{W\left( t\right) }d^p\sigma
\cdot \phi _{,r_2}^{N_2}\cdots \phi _{,r_q}^{N_q}\cdot \Lambda ^{r_2\ldots
r_q}\quad ,
\end{equation}
and due to the first class constraints the summation in the integrand runs
over ''spatial'' indices $r_2,\ldots ,r_q\in \left\{ 1,\ldots ,p\right\} $
only, so that finally%
$$
Y_{MN}=\frac 1{\left( q-1\right) !}S\left( \Delta \right) _{MNN_q\ldots
N_2}\cdot Y_{MN}^{N_2\ldots N_q}\quad , 
$$
and (\ref{ga29}) becomes now 
\begin{equation}
\label{ga31}\left\{ Q_M,Q_N\right\} =\pm C_{MN}^{\;K}\cdot Q_K\;+\;\frac
1{\left( q-1\right) !}S\left( \Delta \right) _{MNN_q\ldots N_2}\cdot
Y_{MN}^{N_2\ldots N_q}\;+Z_{MN}\quad .
\end{equation}

Yet another expression for the above relations can be obtained \cite{Soro}
if we regard the worldvolume as a pseudo-Riemannian manifold with an
(auxiliary) metric which is diagonal in the coordinate system $\left(
t,\sigma \right) $, 
$$
-dt\otimes dt+\delta _{rs}\cdot d\sigma ^r\otimes d\sigma ^s\quad . 
$$
Then the restriction of this metric to the hypersurfaces $W\left( t\right) $
is a Euclidean metric, and $W\left( t\right) $ become Riemannian manifolds,
on which we can introduce a Hodge star operator with respect to this metric.
We need not distinguish between upper and lower indices here, so that $%
\Lambda ^{r_2\ldots r_q}$ for $r_2,\ldots ,r_q\in \left\{ 1,\ldots
,p\right\} $ can be regarded as components of a $\left( q-1\right) $-form $%
\Lambda ^{gauge}$ on $W\left( t\right) $. Its Hodge dual is then 
\begin{equation}
\label{ho1}\left( *\Lambda ^{gauge}\right) _{s_q\ldots s_p}=\frac 1{\left(
q-1\right) !}\,\epsilon _{t_2\ldots t_qs_q\ldots s_p}\cdot \Lambda
^{t_2\ldots t_q}\quad , 
\end{equation}
where all indices are taken from the set $\left\{ 1,\ldots ,p\right\} $. On
using 
\begin{equation}
\label{ho2}\left( **\Lambda ^{gauge}\right) =\left( -1\right) ^{\left(
q-1\right) \left( p-q+1\right) }\cdot \Lambda ^{gauge} 
\end{equation}
we can write%
$$
d\sigma ^1\cdots d\sigma ^p\cdot \phi _{,r_2}^{N_2}\cdots \phi
_{,r_q}^{N_q}\cdot \Lambda ^{r_2\ldots r_q}= 
$$
\begin{equation}
\label{ho3}=\frac 1{\left( p-q+1\right) !}\,\left( *\Lambda ^{gauge}\right)
\cdot emb^{*}d\phi ^{N_2}\cdots emb^{*}d\phi ^{N_q}\quad ; 
\end{equation}
in what follows we shall omit the ''$emb^{*}$'' for the sake of simplicity.
Multiplication of (\ref{ho3}) by $\frac 1{\left( q-1\right) !}S\left( \Delta
\right) _{MNN_q\ldots N_2}$ gives%
$$
d\sigma ^1\cdots d\sigma ^p\cdot S\left( \Delta \right) _{MN}\bullet \Lambda
^{gauge}\;= 
$$
\begin{equation}
\label{ho4}=\;\frac 1{\left( p-q+1\right) !}\,\left( *\Lambda
^{gauge}\right) \cdot S\left( \Delta \right) _{MN}\quad , 
\end{equation}
where $S\left( \Delta \right) _{MN}$ now denotes the pullback of this form
to $W\left( t\right) $, 
\begin{equation}
\label{ho5}S\left( \Delta \right) _{MN}=emb^{*}\,\frac 1{\left( q-1\right)
!}\,d\phi ^{N_2}\cdots d\phi ^{N_q}\cdot S\left( \Delta \right)
_{MNN_q\ldots N_2}\quad . 
\end{equation}
Thus we can rewrite (\ref{ga19}) as 
\begin{equation}
\label{ho6}Y_{MN}\left( t\right) =\frac 1{\left( p-q+1\right)
!}\int\limits_{W\left( t\right) }\,\left( *\Lambda ^{gauge}\right) \cdot
S\left( \Delta \right) _{MN}\quad , 
\end{equation}

We again summarize this section in the form of a theorem.

\subsubsection{Theorem}

Let an Abelian $\left( q-1\right) $-form gauge potential $A_{\mu _2\ldots
\mu _q}$, $q\le p$, be defined on the worldvolume. On the target space a $q$%
-form potential $B$ transforms according to $\delta _MB=d\Delta _M$ under $G$%
. We impose a transformation behaviour $\delta _MA=emb^{*}\Delta _M$ on $A$.
The Lagrangian splits into an invariant piece ${\cal L}_0$ and a
semi-invariant piece ${\cal L}_{WZ}$, as described above. Then

\begin{enumerate}
\item  The Poisson bracket algebra of the {\bf Noether} currents is 
\begin{equation}
\label{th1x1}\left\{ j_M^0,j_N^0\right\} \approx \left[ \pm C_{MN}^{\;K}\cdot
j_K^0\;+\;S\left( \Delta \right) _{MN}\bullet \Lambda ^{gauge}\right] \cdot
\delta \left( \sigma -\sigma ^{\prime }\right) \quad ,
\end{equation}
where $S\left( \Delta \right) _{MN\nu _2\ldots \nu _q}=\left[ emb^{*}S\left(
\Delta \right) _{MN}\right] _{\nu _2\ldots \nu _q}$, and 
\begin{equation}
\label{th2x}S\left( \Delta \right) _{MN}=-\delta _M\Delta _N+\delta _N\Delta
_M\mp C_{MN}^{\;K}\,\Delta _K\;\;
\end{equation}
measures the deviation of the forms $\Delta _M$ from transforming as a
multiplet under the adjoint representation of the group $G$: 
\begin{equation}
\label{th3x}T_M\cdot \Delta _N=\mp \,\Delta _K\cdot ad\left( T_M\right)
_{\;N}^K\;-\;S\left( \Delta \right) _{MN}\quad .
\end{equation}
The once integrated version of (\ref{th1x1}) is 
\begin{equation}
\label{th4x}\left\{ Q_M,j_N^0\right\} =\pm j_K^0\cdot ad\left( T_M\right)
_{\;N}^K\;+\;S\left( \Delta \right) _{MN}\bullet \Lambda ^{gauge}\quad ,
\end{equation}
which defines the action of the generator $T_M$ on the Noether current $j_N^0
$, and $\pm $ refers to a left/right action. The twice integrated version is 
\begin{equation}
\label{th5x}\left\{ Q_M,Q_N\right\} =\pm C_{MN}^{\;K}\cdot
Q_K\;+\;Y_{MN}\quad ,
\end{equation}
where 
\begin{equation}
\label{th6x}Y_{MN}\left( t\right) =\int\limits_{W\left( t\right) }d^p\sigma
\cdot S\left( \Delta \right) _{MN}\bullet \Lambda ^{gauge}\quad .
\end{equation}
If $S\left( \Delta \right) _{MNN_q\ldots N_2}=const.$, then the charges $%
Y_{MN}$ are conserved and central. (\ref{th6x}) can be rewritten in the form 
\begin{equation}
\label{th61}Y_{MN}\left( t\right) =\frac 1{\left( p-q+1\right)
!}\int\limits_{W\left( t\right) }\,\left( *\Lambda ^{gauge}\right) \cdot
S\left( \Delta \right) _{MN}\quad ,
\end{equation}
with $*\Lambda ^{gauge}$ being the Hodge dual of the form $\Lambda
^{gauge}=\frac 1{\left( q-1\right) !}d\sigma ^{r_2}\cdots d\sigma
^{r_q}\cdot \Lambda ^{r_2\ldots r_q}$ on the worldvolume.

\item  The Poisson bracket algebra of the {\bf modified} currents is 
\begin{equation}
\label{th7x}\left\{ \widetilde{j_M^0},\widetilde{j_N^0}\right\} \approx \pm
C_{MN}^{\;K}\cdot \widetilde{j_K^0}\cdot \delta \left( \sigma -\sigma
^{\prime }\right) \;+\;\left[ S\left( U\right) _{MN}\;+\;S\left( \Delta
\right) _{MN}\bullet \Lambda ^{gauge}\;\right] \quad ,
\end{equation}
with 
\begin{equation}
\label{th8x}S\left( U\right) _{MN}=\delta _M\phi ^K\cdot \frac{\partial U_N^0%
}{\partial \phi ^K}-\delta _N\phi ^K\cdot \frac{\partial U_M^0}{\partial
\phi ^K}\pm C_{MN}^{\;K}\cdot U_K^0
\end{equation}
and its integral 
\begin{equation}
\label{th9x}Z_{MN}=\int\limits_{W\left( t\right) }d^p\sigma \cdot S\left(
U\right) _{MN}\quad .
\end{equation}
Double integration of the current algebra yields the charge algebra 
\begin{equation}
\label{th10x}\left\{ Q_M,Q_N\right\} =\pm C_{MN}^{\;K}\cdot
Q_K\;+\;Y_{MN}\;+\;Z_{MN}\quad .
\end{equation}
In the case of constant $S\left( \Delta \right) _{MN}$ this can be written
as 
\begin{equation}
\label{th11x}\left\{ Q_M,Q_N\right\} =C_{MN}^{\;K}\cdot Q_K\;+\;\frac
1{\left( q-1\right) !}\,S\left( \Delta \right) _{MNN_q\ldots N_2}\cdot
Y_{MN}^{N_2\ldots N_q}\;+Z_{MN}\quad .
\end{equation}
\end{enumerate}

\section{Extended superalgebras carried by $D$-$p$-branes in IIA superspace 
\label{SuperAlgebra}}

Now we apply the ideas we have developed in the previous sections to the
case of $D$-$p$-branes in IIA superspace. This restricts $p$ to be even, $%
p=0,2,4,6,8$. However, before doing so, we first discuss our superspace
conventions, and our definitions of graded Poisson brackets. Then we first
compute the algebra of Noether charges and of modified Noether charges
resulting from a D-brane Lagrangian without specific assumptions on the
background the brane propagates in, or on the specific form of the various
gauge fields occuring in the Lagrangian. Then we recapitulate how
supergravity determines the background in which the branes propagate, and
the relation of superspace constraints with $\kappa $-symmetry of the
branes. Then we study the Bianchi identities associated with a specific
choice of background gauge fields in superspace; and only then we work out
the explicit superalgebra extensions carried by D-branes in this particular $%
D=10$ vacuum.

\subsection{Conventions}

\subsubsection{Superspace conventions \label{conventions}}

The target space $\Sigma $ is now the coset space%
$$
\mbox{IIA-superMinkowski}=\mbox{IIA-superPoincare}/SO\left( 1,9\right)  
$$
with coordinates $\left( X,\theta \right) $ that label the coset
representative $e^{iX\cdot P+\theta Q}$. Adopting the convention that the
complex conjugate of a product of two spinors reverses their order this
implies that in an operator realization the coset representatives are mapped
to unitary operators, provided that $P$ and $Q$ are hermitian. The
assumption of IIA superspace means that we have two $16$-component spinor
generators of opposite chirality which transform under the two irreducible $%
\left( 16\times 16\right) $-dimensional spin representations of $SO\left(
1,9\right) $; but effectively, this yields one non-chiral $32$-component
spinor, transforming under the direct sum of the two irreducible spin
representations, which is just the representation of $SO\left( 1,9\right) $
obtained from the $32$-component $\Gamma $-matrices. The metric $\eta _{mn}$
on the target space is flat $10$-dimensional ''mostly plus'' Minkowski
metric. Spinor components occur with natural index up; an inner product
between spinors is provided by the bilinear form $\left( \chi ,\theta
\right) \mapsto \chi ^\alpha C_{\alpha \beta }\theta ^\beta $, where $C$ is
a charge conjugation matrix. In $D=10$ and with the Minkowski metric as
specified above we can choose a Majorana-Weyl representation for the spinors
and the $\Gamma $-matrices, respectively, in which spinors have real
Grassmann-odd components, the matrices $C\Gamma _m$ are real and symmetric,
and $C$ is real and antisymmetric. In such a representation we can choose $%
C=\pm \Gamma _0$. Altogether we have $32$ real fermion degrees of freedom a
priori. Spinor indices are lowered and raised {\bf from the left} with the
charge conjugation matrix and its inverse, respectively; e.g., raising is
accomplished with the inverse of $C$, the components of which are denoted by 
$C^{\alpha \beta }$, by $\theta _\beta \mapsto \theta ^\alpha =C^{\alpha
\beta }\theta _\beta $. By definition, $C^{\alpha \beta }C_{\beta \gamma
}=\delta _\gamma ^\alpha $. An expression like $\bar \epsilon \Gamma
_m\theta $ therefore means%
$$
\bar \epsilon \Gamma _m\theta =\epsilon ^\alpha C_{\alpha \beta }\left(
\Gamma _m\right) _{\;\gamma }^\beta \theta ^\gamma \quad , 
$$
etc. Our supertranslation algebra is 
\begin{equation}
\label{ap1}\left\{ Q_\alpha ,Q_\beta \right\} =2\Gamma _{\alpha \beta
}^m\cdot P_m\quad .
\end{equation}
The action of $e^{iY\cdot P+\epsilon Q}$ on $\left( X,\theta \right) $
yields $\left( X^{\prime },\theta ^{\prime }\right) $, where $\left(
X^{\prime },\theta ^{\prime }\right) $ is implicitly defined by 
\begin{equation}
\label{ap2}e^{iY\cdot P+\epsilon Q}e^{iX\cdot P+\theta Q}=e^{iX^{\prime
}\cdot P+\theta ^{\prime }Q}\quad ;
\end{equation}
for infinitesimal $\epsilon $ this yields $\left( X^{\prime },\theta
^{\prime }\right) =\left( X+Y+i\bar \epsilon \Gamma \theta ,\theta +\epsilon
\right) $. From (\ref{ap2}) it can be seen that this is a left action. The
vector fields $\widetilde{T_\alpha }$, $i\widetilde{T_m}$ induced by the
generators $Q_\alpha $, $iP_m$ on $\Sigma $ are therefore 
\begin{equation}
\label{ap3}\widetilde{T_\alpha }=\left( i\Gamma ^m\theta \right) _\alpha
\cdot \frac \partial {\partial X^m}+\frac \partial {\partial \theta ^\alpha
}=:\delta _\alpha \quad ,
\end{equation}
\begin{equation}
\label{ap4}i\widetilde{T_m}=\frac \partial {\partial X^m}=:\delta _m\quad .
\end{equation}
By construction they are right-invariant vector fields. The corresponding
left-invariant vector fields are obtained by replacing $\theta \mapsto
-\theta $ in (\ref{ap3}, \ref{ap4}). Their duals are the left invariant $1$%
-forms $\Pi ^M=\left( \Pi ^m,\Pi ^a\right) $ on superspace, where
\begin{equation}
\label{ap44}\Pi ^m=dX^m+id\bar \theta \Gamma ^m\theta \quad ,\quad \Pi
^\alpha =d\theta ^\alpha \quad .
\end{equation}
From (\ref{ap4}) we see that $\delta _m$ is strictly speaking ''$i\times $
Poincare-translation with generator $P_m$''.

The graded Lie-bracket of $\widetilde{T_\alpha }$, $\widetilde{T_\beta }$ is 
$\left\{ \delta _\alpha ,\delta _\beta \right\} =\left[ \widetilde{T_\alpha }%
,\widetilde{T_\beta }\right] _{graded\;Lie}=$%
\begin{equation}
\label{ap5}=\left( -2\Gamma _{\alpha \beta }^m\right) \cdot \left( -i\frac
\partial {\partial X^m}\right) =-2\Gamma _{\alpha \beta }^m\cdot \widetilde{%
T_m}=2i\Gamma _{\alpha \beta }^m\cdot \delta _m\quad ,
\end{equation}
i.e. the algebra (\ref{ap1}) is satisfied up to a sign, which is in accord
with (\ref{zw2}) in the first section, since the action of the supergroup on 
$\Sigma $ is from the left, or equivalently, since the $\widetilde{T_M}$ are
right-invariant.

Summation of superspace indices is defined according to%
$$
\omega =\frac 1{r!}dZ^{M_1}\ldots dZ^{M_r}\cdot \omega _{M_r\ldots M_1}\quad
, 
$$
where $\omega $ is a superspace $p$-form. The forms on the worldvolume obey
the usual summation conventions, however; e.g. for the pull-back of the
above $r$-form to the worldvolume we write%
$$
emb^{*}\omega =\frac 1{r!}\partial _{\mu _1}Z^{M_1}\ldots \partial _{\mu
_r}Z^{M_r}\cdot \omega _{M_r\ldots M_1}\cdot dx^{\mu _1}\ldots dx^{\mu
_r}\quad . 
$$
Exterior derivative $d$ is defined to act from the right on superspace forms
as well as on worldvolume forms,%
$$
d\left( \omega \chi \right) =\omega d\chi +\left( -1\right) ^qd\omega \cdot
\chi \quad , 
$$
where $\chi $ is a $q$-form.

\subsubsection{Graded Poisson brackets}

Given two (possibly graded) functionals $F$, $G$ of the (possibly graded)
time-dependent fields $\phi ^i\left( t,\sigma \right) $ and their canonical
conjugate momenta $\Lambda _i\left( t,\sigma \right) $ that are defined on a 
$p$-dimensional manifold $S$ with coordinates $\left( \sigma ^1,\ldots
,\sigma ^p\right) $, their Poisson bracket is defined by (see, e.g., \cite
{Govaerts}) 
\begin{equation}
\label{ap6}\left\{ F,G\right\} _{PB}=\int\limits_Sd^p\sigma \,\sum_i\left[
\left( -1\right) ^{\phi ^i}F\frac{\overleftarrow{\delta }}{\delta \phi
^i\left( \sigma \right) }\frac{\overrightarrow{\delta }}{\delta \Lambda
_i\left( \sigma \right) }G-F\frac{\overleftarrow{\delta }}{\delta \Lambda
_i\left( \sigma \right) }\frac{\overrightarrow{\delta }}{\delta \phi
^i\left( \sigma \right) }G\right] \quad ; 
\end{equation}
this can be expressed in terms of derivatives acting solely from the left by 
\begin{equation}
\label{ap7}\left\{ F,G\right\} _{PB}=\int\limits_Sd^p\sigma \,\sum_i\left[
\left( -1\right) ^{F\phi ^i}\frac{\delta F}{\delta \phi ^i\left( \sigma
\right) }\frac{\delta G}{\delta \Lambda _i\left( \sigma \right) }-\left(
-1\right) ^{\left( F+1\right) \phi ^i}\frac{\delta F}{\delta \Lambda
_i\left( \sigma \right) }\frac{\delta G}{\delta \phi ^i\left( \sigma \right) 
}\right] \quad , 
\end{equation}
where $\left( -1\right) ^{F\phi ^i}=1$ {\bf iff} both $F$ and $\phi ^i$ are
Grassmann-odd. With these definitions the following rules are satisfied:

\begin{enumerate}
\item  Graded Antisymmetry, 
\begin{equation}
\label{ap8}\left\{ F,G\right\} =-\left( -1\right) ^{FG}\left\{ G,F\right\}
\quad .
\end{equation}

\item  Graded Leibnitz rule, 
\begin{equation}
\label{ap9}\left\{ F,GH\right\} =\left\{ F,G\right\} H+\left( -1\right)
^{FG}G\left\{ F,H\right\} \quad .
\end{equation}

\item  Graded Jacobi identity, 
\begin{equation}
\label{ap10}\left( -1\right) ^{FH}\left\{ F,\left\{ G,H\right\} \right\}
+\left( -1\right) ^{GF}\left\{ G,\left\{ H,F\right\} \right\} +\left(
-1\right) ^{HG}\left\{ H,\left\{ F,G\right\} \right\} =0\quad .
\end{equation}
\end{enumerate}

\subsection{D-$p$ brane Lagrangians}

\subsubsection{Structure of the Lagrangian}

The kinetic supertranslation-invariant part ${\cal L}_0$ in the D-$p$-brane
Lagrangian is given by 
\begin{equation}
\label{an1}{\cal L}_0=\sqrt{-\det \left( g_{\mu \nu }+\widehat{F}_{\mu \nu
}\right) }\quad ,
\end{equation}
where $g_{\mu \nu }=\Pi _\mu ^m\Pi _\nu ^n\eta _{nm}$ is the pull-back of
the $10$-dimensional ''mostly plus'' Minkowski metric $\eta _{nm}$ to the
worldvolume $W$ of the D-$p$-brane using left-invariant (LI) $1$-forms $\Pi
^A$, and 
\begin{equation}
\label{an2}\widehat{F}_{\mu \nu }=\partial _\mu A_\nu -\partial _\nu A_\mu
-\Pi _\mu ^{A_1}\Pi _\nu ^{A_2}\cdot B_{A_2A_1}\quad ,
\end{equation}
where $F_{\mu \nu }:=\partial _\mu A_\nu -\partial _\nu A_\mu $ are the
components of the field strength $F$ of the gauge potential $A$ defined on
the worldvolume, and $B_{A_2A_1}$ are the components of the superspace $2$%
-form potential $B$ in the LI-basis whose leading component in a $\theta $%
-expansion is the NS-NS gauge potential. In the discussion below we shall
assume that its bosonic components are zero, but at present $B$ could be
quite arbitrary. Under supertranslations $\delta _\alpha $ the field
strength $H=dB$ is assumed to be invariant,
\begin{equation}
\label{an20}\delta _\alpha H=0\quad .
\end{equation}
This implies that $B$ transforms locally as a differential, 
\begin{equation}
\label{an3}\quad \delta _\alpha B=d\Delta _\alpha \quad .
\end{equation}
Therefore it transforms as a differential under Poincare translations as
well: To see this compare%
$$
2i\Gamma _{\alpha \beta }^m\cdot \delta _mB=\left\{ \delta _\alpha ,\delta
_\beta \right\} B=d\left( \delta _\alpha \Delta _\beta +\delta _\beta \Delta
_\alpha \right) \quad , 
$$
where in the first equation (\ref{ap5}) has been used. If we multiply with
another $\Gamma $-matrix and take the trace we find that
\begin{equation}
\label{an310}\delta _mB=d\Delta _m
\end{equation}
with
\begin{equation}
\label{an4}\Delta _m=\frac{\Gamma _m^{\alpha \beta }}{i\cdot tr\left( {\bf 1}%
_{32}\right) }\,\delta _\alpha \Delta _\beta \quad .
\end{equation}
Since $\left[ \delta _m,\delta _n\right] =0$ it follows from (\ref{an310})
that $d\left( \delta _m\Delta _n-\delta _n\Delta _m\right) =0$, which
implies, that locally 
\begin{equation}
\label{an5}\delta _m\Delta _n-\delta _n\Delta _m=df_{mn}
\end{equation}
for some function $f_{mn}$. This function need not be defined globally,
however.

Although $A_{\mu \,}$ is a worldvolume field it is defined to transform
under these translations according to 
\begin{equation}
\label{an6}\delta _mA=emb^{*}\Delta _m\quad ,\quad \delta _\alpha
A=emb^{*}\Delta _\alpha \quad , 
\end{equation}
where $emb:W\rightarrow \Sigma $ denotes the embedding of the worldvolume
into the target space, and $emb^{*}$ denotes the associated pull-back. With
this definition the quantity $\widehat{F}$ is invariant under Poincare- and
supertranslations, as explained in section \ref{Eich}. Since the same is
true for $g_{\mu \nu }$ we see that therefore ${\cal L}_0$ is invariant as
well.

\subsubsection{Wess-Zumino term}

The Wess-Zumino form $\left( WZ\right) $ in the D-$p$-brane Lagrangian is
the $\left( p+1\right) $-form on the worldvolume 
\begin{equation}
\label{an7}\left( WZ\right) =\sum\limits_{n=0}^{\left[ \frac{p+1}2\right]
}\frac 1{n!}\,emb^{*}C^{\left( p+1-2n\right) }\cdot \widehat{F}^n\quad ,
\end{equation}
where $C^{\left( r\right) }$ are the superspace potentials whose leading
components in a $\theta $-expansion are the usual bosonic RR gauge
potentials \cite{BergTown1}; in the discussion below the bosonic components
of its field strengths will be set to zero, which then amounts to the choice
of a particular background, but here we make no specific assumptions on the
form of $C^{\left( r\right) }$. The pull-back of $\left( WZ\right) $ to the
worldvolume gives the Wess-Zumino term ${\cal L}_{WZ}$ in the Lagrangian, 
\begin{equation}
\label{an8}{\cal L}_{WZ}=\sum\limits_{n=0}^{\left[ \frac{p+1}2\right] }\frac{%
\epsilon ^{\lambda _1\ldots \lambda _{p+1-2n}\nu _1\ldots \nu _{2n}}}{\left(
p+1-2n\right) !\cdot 2^n\cdot n!}\,\left[ emb^{*}C^{\left( p+1-2n\right)
}\right] _{\lambda _1\ldots \lambda _{p+1-2n}}\widehat{F}_{\nu _1\nu
_2}\cdots \widehat{F}_{\nu _{2n-1}\nu _{2n}}\quad .
\end{equation}
Since we are in IIA superspace we have actually $p=2q$, $q=0,\ldots ,4$.

If $\delta $ denotes either a supertranslation or a Poincare translation
then invariance of $\widehat{F}$ under either of these transformations
implies that 
\begin{equation}
\label{lag1}\delta {\cal L}_{WZ}=\sum\limits_{n=0}^{\left[ \frac{p+1}%
2\right] }\frac{\epsilon ^{\lambda _1\ldots \lambda _{p+1-2n}\nu _1\ldots
\nu _{2n}}}{\left( p+1-2n\right) !\cdot 2^n\cdot n!}\,\left[ emb^{*}\delta
C^{\left( p+1-2n\right) }\right] _{\lambda _1\ldots \lambda _{p+1-2n}}%
\widehat{F}_{\nu _1\nu _2}\cdots \widehat{F}_{\nu _{2n-1}\nu _{2n}}\quad .
\end{equation}
The field strengths associated with the RR-potentials $C^{\left( r\right) }$
are defined to be 
\begin{equation}
\label{lag2}R^{\left( r+1\right) }=\left\{ 
\begin{array}{ccc}
dC^{\left( r\right) } & ; & r=0,1\; \\ 
dC^{\left( r\right) }-C^{\left( r-2\right) }H & ; & r=2,\ldots ,10
\end{array}
\right. \quad ;
\end{equation}
they obey the Bianchi identities 
\begin{equation}
\label{lag3}\left\{ 
\begin{array}{ccc}
dR^{\left( r+1\right) }=0 & ; & r=0,1\; \\ 
dR^{\left( r+1\right) }-R^{\left( r-1\right) }H=0 & ; & r=2,\ldots ,10
\end{array}
\right. \quad .
\end{equation}
It is now assumed that the field strengths (\ref{lag2}) are supertranslation
invariant, 
\begin{equation}
\label{lag4}\delta _\alpha R^{\left( r+1\right) }=0\quad ;\quad r=0,\ldots
,10\quad .
\end{equation}
Then it follows from 
\begin{equation}
\label{lag5}\delta _m=\frac{\Gamma _m^{\alpha \beta }}{2i\cdot tr\left( {\bf %
1}_{32}\right) }\cdot \left\{ \delta _\alpha ,\delta _\beta \right\} 
\end{equation}
that it is invariant under Poincare-translations $-i\delta _m$ as well. From
(\ref{an20}, \ref{lag3}, \ref{lag4}) we can construct the general form of $%
\delta _\alpha C$ for both the IIA and IIB case recursively by starting with
the lowest rank form $C^{\left( 1\right) }$ or $C^{\left( 0\right) }$,
respectively. For the IIA case the result is that there exist superspace
forms 
\begin{equation}
\label{lag6}D_\alpha ^{\left( 2r\right) }\quad ;\quad r=0,\ldots ,4\quad
;\quad \alpha =1,\ldots ,32
\end{equation}
such that 
\begin{equation}
\label{lag7}\delta _\alpha C^{\left( 2q+1\right)
}=\sum\limits_{k=0}^qdD_\alpha ^{\left( 2q-2k\right) }\cdot \frac{B^k}{k!}%
\quad .
\end{equation}
The superscript $\left( 2r\right) $ in (\ref{lag6}) refers to the fact that
the index $\alpha $ does not take part in a summation, but labels one of $32$
components of a spinor-valued $\left( 2r\right) $-form $D^{\left( 2r\right)
}=\left( D_\alpha ^{\left( 2r\right) }\right) _{\alpha =1,\ldots ,32}$.
Using (\ref{lag5}) we find that 
\begin{equation}
\label{lag8}\delta _mC^{\left( 2q+1\right) }=\sum\limits_{k=0}^qdD_m^{\left(
2q-2k\right) }\cdot \frac{B^k}{k!}\quad ,
\end{equation}
where 
\begin{equation}
\label{lag9}D_m^{\left( 0\right) }=\frac{\Gamma _m^{\alpha \beta }}{2i\cdot
tr\left( {\bf 1}_{32}\right) }\cdot \left[ \delta _\alpha D_\beta ^{\left(
0\right) }+\delta _\beta D_\alpha ^{\left( 0\right) }\right] \;=\;\frac{%
\Gamma _m^{\alpha \beta }}{i\cdot tr\left( {\bf 1}_{32}\right) }\cdot \delta
_\alpha D_\beta ^{\left( 0\right) }\quad ,
\end{equation}
$$
D_m^{\left( 2q\right) }=\frac{\Gamma _m^{\alpha \beta }}{2i\cdot tr\left( 
{\bf 1}_{32}\right) }\cdot \left[ \delta _\alpha D_\beta ^{\left( 2q\right)
}+\delta _\beta D_\alpha ^{\left( 2q\right) }+D_\alpha ^{\left( 2q-2\right)
}\cdot d\Delta _\beta +D_\beta ^{\left( 2q-2\right) }\cdot d\Delta _\alpha
\right] \;= 
$$
\begin{equation}
\label{lag10}=\;\frac{\Gamma _m^{\alpha \beta }}{i\cdot tr\left( {\bf 1}%
_{32}\right) }\cdot \left[ \delta _\alpha D_\beta ^{\left( 2q\right)
}+D_\alpha ^{\left( 2q-2\right) }\cdot d\Delta _\beta \right] \quad .
\end{equation}

Now let us return to the supertranslation variation of $C^{\left( r\right) }$
in (\ref{lag7}). If we insert (\ref{lag7}) in (\ref{lag1}) we find that all
terms involving $B$ cancel, 
\begin{equation}
\label{lag11}\delta _\alpha {\cal L}_{WZ}=\partial _\mu U_\alpha ^\mu \quad
, 
\end{equation}
with 
\begin{equation}
\label{lag12}U_\alpha ^\mu =\sum\limits_{n=0}^q\frac{\epsilon ^{\mu \mu
_2\ldots \mu _{2q+1-2n}\nu _1\ldots \nu _{2n}}}{\left( 2q-2n\right) !\cdot
2^n\cdot n!}\,\left[ emb^{*}D_\alpha ^{\left( 2q-2n\right) }\right] _{\mu
_2\ldots \mu _{2q+1-2n}}F_{\nu _1\nu _2}\cdots F_{\nu _{2n-1}\nu _{2n}}\quad
. 
\end{equation}
Similarly, application of (\ref{lag5}) gives 
\begin{equation}
\label{lag13}U_m^\mu =-i\sum\limits_{n=0}^q\frac{\epsilon ^{\mu \mu _2\ldots
\mu _{2q+1-2n}\nu _1\ldots \nu _{2n}}}{\left( 2q-2n\right) !\cdot 2^n\cdot n!%
}\,\left[ emb^{*}D_m^{\left( 2q-2n\right) }\right] _{\mu _2\ldots \mu
_{2q+1-2n}}F_{\nu _1\nu _2}\cdots F_{\nu _{2n-1}\nu _{2n}}\quad , 
\end{equation}
with the $D_m$ given in (\ref{lag9}, \ref{lag10}). We observe that $U_\alpha
^0,U_m^0$ contain neither of $\dot X,\dot \theta ,F_{0r}$.

\subsubsection{Noether currents and Noether charges}

We now want to compute the algebra of Noether currents and Noether charges
resulting from these currents; as discussed in section \ref{Eich} we may
expect that due to the fact that the worldvolume gauge field $A_\mu \,$
takes part in the supersymmetry transformations on the target space even the
current algebra of the {\bf Noether} currents fails to close in the ordinary
form, but will be extended by central pieces. All the more this will be true
for the algebra of the modified currents and charges, respectively.

Our degrees of freedom are now $\left( X^m,\theta ^\alpha ,A_\nu \right) $
with $\nu =0,\ldots ,p$; the associated canonical conjugate momenta are $%
\left( \Lambda _m,\Lambda _\alpha ,\Lambda ^\nu \right) $, respectively. As
discussed in section \ref{Eich} we have a primary constraint $\Lambda ^0=0$
which amounts to a reduction of phase space, and a secondary constraint $%
\sum_{r=1}^p\partial _r\Lambda ^r=0$, which holds only on-shell, i.e. on
using the equations of motion. The zeroth components of the Noether currents
associated with the generators $Q_\alpha $ are 
\begin{equation}
\label{an9}j_\alpha ^0=\left( i\Gamma ^m\theta \right) _\alpha \cdot \Lambda
_m+\Lambda _\alpha +\left( emb^{*}\Delta _\alpha \right) _\nu \cdot \Lambda
^\nu \quad ; 
\end{equation}
as explained in section \ref{Eich} the primary constraint $\Lambda ^0=0$
must not be taken into account before all Poisson brackets have been worked
out. The zeroth components of the Noether currents associated with the
generators $P_m$ are 
\begin{equation}
\label{an10}j_m^0=-i\Lambda _m-i\left( emb^{*}\Delta _m\right) _\nu \cdot
\Lambda ^\nu \quad . 
\end{equation}
Then the Poisson bracket of the currents $\left\{ j_\alpha ^0,j_\beta
^0\right\} $ is%
$$
\left\{ j_\alpha ^0\left( t,\sigma \right) ,j_\beta ^0\left( t,\sigma
^{\prime }\right) \right\} \approx -2i\Gamma _{\alpha \beta }^m\cdot \Lambda
_m\cdot \delta \left( \sigma -\sigma ^{\prime }\right) \;- 
$$
\begin{equation}
\label{an11}-\;\left[ \delta _\alpha \left( emb^{*}\Delta _\beta \right)
_r+\delta _\beta \left( emb^{*}\Delta _\alpha \right) _r\right] \cdot
\Lambda ^r\cdot \delta \left( \sigma -\sigma ^{\prime }\right) \quad , 
\end{equation}
where $"\approx "$ means ''on using all constraints and equations of motion
and on discarding surface terms''. If we insert (\ref{an10}) into the last
equation we get 
\begin{equation}
\label{an12}\left\{ j_\alpha ^0,j_\beta ^0\right\} \approx \left[ 2\Gamma
_{\alpha \beta }^m\cdot j_m^0\;+\left( emb^{*}S_{\alpha \beta }\left( \Delta
\right) \right) _r\cdot \Lambda ^r\right] \cdot \delta \left( \sigma -\sigma
^{\prime }\right) \quad , 
\end{equation}
where we have used the primary constraint $\Lambda ^0=0$ at last. $S_{\alpha
\beta }\left( \Delta \right) $ is given by 
\begin{equation}
\label{an13}S_{\alpha \beta }\left( \Delta \right) =2i\Gamma _{\alpha \beta
}^n\cdot \Delta _m-\delta _\alpha \Delta _\beta -\delta _\beta \Delta
_\alpha \quad . 
\end{equation}
We see that the presence of the gauge field $A_\mu $ taking part in the
supersymmetry variation of the Lagrangian alters the form of the algebra
even of the {\bf Noether} currents, i.e. before taking into account the
possible modifications of the Noether currents by terms originating in the
Wess-Zumino term.

The once integrated version of (\ref{an12}) defines the action of the
generator $Q_\alpha $ on the current $j_\beta ^0$, 
\begin{equation}
\label{an14}\left\{ Q_\alpha ,j_\beta ^0\right\} =2\Gamma _{\alpha \beta
}^m\cdot j_m^0\;+\left( emb^{*}S_{\alpha \beta }\left( \Delta \right)
\right) _r\cdot \Lambda ^r\quad . 
\end{equation}
As explained in section \ref{NoethChar} the presence of the Wess-Zumino term
in the Lagrangian implies that the Noether charges $Q_\alpha $ which are
obtained by integrating the zero components $j_\alpha ^0$ over the
hypersurface $W\left( t\right) $ are no longer conserved; however, if the
Wess-Zumino term ${\cal L}_{WZ}$ and the NS-NS gauge potential are
translational invariant, as will be the case below, the Noether charges $P_m$
obtained by integrating $j_m^0$ are still conserved.

The twice integrated version of (\ref{an12}) describes the modified algebra
of the Noether charges, 
\begin{equation}
\label{an15}\left\{ Q_\alpha ,Q_\beta \right\} =2\Gamma _{\alpha \beta
}^m\cdot P_m+\int\limits_{W\left( t\right) }d^p\sigma
\sum\limits_{r=1}^p\left( emb^{*}S_{\alpha \beta }\left( \Delta \right)
\right) _r\cdot \Lambda ^r\quad ;
\end{equation}
here $\left( emb^{*}S_{\alpha \beta }\left( \Delta \right) \right) _r=$ $%
\partial _rZ^M\cdot S_{\alpha \beta M}$, when $S_{\alpha \beta }$ is
expanded in the coordinate basis $\left( dZ^M\right) =\left( dX^m,d\theta
^\alpha \right) $. Now let us assume (see section \ref{Eich}) that $%
S_{\alpha \beta M}$ are constants; we want to find extensions of the current
and charge algebra by topological charges carried by the brane; but since it
is only the bosonic coordinates $X^m$ and the pull-back of their
differentials to the worldvolume that describe the topology of the image $%
embW\left( t\right) $ of the brane in the spacetime $\Sigma $ we need only
consider the terms involving bosonic $1$-forms $dX^m$, i.e. $\partial _rX^m$%
, in the above pull-back; therefore if we now define the charge 
\begin{equation}
\label{an16}Y^m=\int\limits_{W\left( t\right) }d^p\sigma
\sum\limits_{r=1}^p\partial _rX^m\cdot \Lambda ^r\quad ,
\end{equation}
then the algebra of the Noether charges in (\ref{an15}) becomes 
\begin{equation}
\label{an17x}\left\{ Q_\alpha ,Q_\beta \right\} =2\Gamma _{\alpha \beta
}^m\cdot P_m+S_{\alpha \beta m}\cdot Y^m\quad ,\quad S_{\alpha \beta m}\quad 
\mbox{constant.}
\end{equation}
If we think of $W\left( t\right) $ as being endowed with an auxiliary
Euclidean metric which is diagonal in the coordinates $\left( \sigma
^r\right) $ then we can introduce the Hodge dual of the $1$-form $\Lambda
^{gauge}=\sum_{r=1}^pd\sigma ^r\Lambda ^r$ \cite{Soro}, which is given by 
\begin{equation}
\label{an18x}\left( *\Lambda ^{gauge}\right) _{s_2\ldots s_p}=\epsilon
_{rs_2\ldots s_p}\Lambda ^r\quad ,
\end{equation}
and rewrite (\ref{an16}) as 
\begin{equation}
\label{an17}Y^m=\frac 1{\left( p-1\right) !}\int\limits_{W\left( t\right)
}\left( *\Lambda ^{gauge}\right) \,dX^m\quad ,
\end{equation}
where $dX^m$ now denotes the pull-back $emb^{*}dX^m$, and exterior product
of forms is understood in the integrand. Furthermore we note that $*\Lambda
^{gauge}$ is {\bf closed} on the physical trajectories, since 
\begin{equation}
\label{an171}d*\Lambda ^{gauge}=\partial _r\Lambda ^r\cdot d\sigma ^1\cdots
d\sigma ^p\quad .
\end{equation}

A similar computation now shows that 
\begin{equation}
\label{an18}\left\{ j_\alpha ^0\left( t,\sigma \right) ,j_m^0\left( t,\sigma
^{\prime }\right) \right\} \approx \left( emb^{*}S_{\alpha m}\left( \Delta
\right) \right) _r\cdot \Lambda ^r\cdot \delta \left( \sigma -\sigma
^{\prime }\right) \quad ,
\end{equation}
with 
\begin{equation}
\label{an19}S_{\alpha m}\left( \Delta \right) =i\left( \delta _\alpha \Delta
_m-\delta _m\Delta _\alpha \right) \quad ;
\end{equation}
the factor of $i$ comes from our parametrising of the coset elements of
superMinkowski space, see section \ref{conventions}.

Finally, we find 
\begin{equation}
\label{an20x}\left\{ j_m^0\left( t,\sigma \right) ,j_n^0\left( t,\sigma
^{\prime }\right) \right\} \approx \left( emb^{*}S_{mn}\left( \Delta \right)
\right) _r\cdot \Lambda ^r\cdot \delta \left( \sigma -\sigma ^{\prime
}\right) \quad ,
\end{equation}
where 
\begin{equation}
\label{an21}S_{mn}\left( \Delta \right) =\delta _m\Delta _n-\delta _n\Delta
_m\quad .
\end{equation}
Double integration of (\ref{an20x}) using (\ref{an5}) then yields 
\begin{equation}
\label{an22}\left[ P_m,P_n\right] =\int\limits_{W\left( t\right) }d^p\sigma
\cdot \partial _r\left( f_{mn}\Lambda ^r\right) \quad ,
\end{equation}
where we have used the secondary constraint $\partial _r\Lambda ^r=0$ and
the equations of motion. We see that the momenta can be {\bf non-commuting}
in the case that the functions $f_{mn}$ are not globally defined; this could
happen if some of the dimensions of $W\left( t\right) $ are compact, and
their images in the spacetime under the embedding describe a closed but
non-contractible cycle.

\subsubsection{Modified currents and charges \label{Modificatio}}

The modified currents are 
\begin{equation}
\label{an23}\widetilde{j_\alpha ^0}=j_\alpha ^0-U_\alpha ^0\quad ,\quad 
\widetilde{j_m^0}=j_m^0-U_m^0\quad . 
\end{equation}
Their Poisson brackets are found to be 
\begin{equation}
\label{an24}\left\{ \widetilde{j_\alpha ^0},\widetilde{j_\beta ^0}\right\}
\approx \left[ 2\Gamma _{\alpha \beta }^n\cdot \widetilde{j_m^0}+\left(
emb^{*}S_{\alpha \beta }\left( \Delta \right) \right) _r\cdot \Lambda
^r+S_{\alpha \beta }\left( U\right) \right] \cdot \delta \left( \sigma
-\sigma ^{\prime }\right) \quad , 
\end{equation}
with 
\begin{equation}
\label{an25}S_{\alpha \beta }\left( U\right) =\delta _\alpha U_\beta
^0+\delta _\beta U_\alpha ^0+2\Gamma _{\alpha \beta }^n\cdot U_n^0\quad . 
\end{equation}
$S_{\alpha \beta }$ is given in (\ref{an13}). Furthermore, 
\begin{equation}
\label{an26}\left\{ \widetilde{j_\alpha ^0},\widetilde{j_m^0}\right\}
\approx \left[ \left( emb^{*}S_{\alpha m}\left( \Delta \right) \right)
_r\cdot \Lambda ^r+S_{\alpha m}\left( U\right) \right] \cdot \delta \left(
\sigma -\sigma ^{\prime }\right) \quad , 
\end{equation}
\begin{equation}
\label{an27}S_{\alpha m}\left( U\right) =\delta _\alpha U_m^0+i\delta
_mU_\alpha ^0\quad , 
\end{equation}
with $S_{\alpha m}$ given in (\ref{an19}); and finally, 
\begin{equation}
\label{an28}\left\{ \widetilde{j_m^0},\widetilde{j_n^0}\right\} \approx
\left[ \left( emb^{*}S_{mn}\left( \Delta \right) \right) _r\cdot \Lambda
^r+S_{mn}\left( U\right) \right] \cdot \delta \left( \sigma -\sigma ^{\prime
}\right) \quad , 
\end{equation}
\begin{equation}
\label{an29x}S_{mn}\left( U\right) =-i\left[ \delta _mU_n^0-\delta
_nU_m^0\right] \quad , 
\end{equation}
with $S_{mn}$ from (\ref{an21}).

We can work out the expressions for $S_{MN}\left( U\right) $ using (\ref
{lag12}, \ref{lag13}), which yields%
$$
S_{\alpha \beta }\left( U\right) \;=\;\frac{\epsilon ^{0\nu _1\ldots \nu
_{2q}}}{2^q\cdot q!}\,emb^{*}\left[ \delta _\alpha D_\beta ^{\left( 0\right)
}+\delta _\beta D_\alpha ^{\left( 0\right) }+2\Gamma _{\alpha \beta }^n\cdot
D_n^{\left( 0\right) }\right] \cdot F_{\nu _1\nu _2}\cdots F_{\nu _{2q-1}\nu
_{2q}}\;+ 
$$
$$
+\;\sum\limits_{k=0}^{q-1}\frac{\epsilon ^{0\mu _2\ldots \mu _{2q+1-2k}\nu
_1\ldots \nu _{2k}}}{\left( 2q-2k\right) !\,2^k\cdot k!}\,\left\{
emb^{*}\left[ \delta _\alpha D_\beta ^{\left( 2q-2k\right) }+\delta _\beta
D_\alpha ^{\left( 2q-2k\right) }+\right. \right. 
$$
$$
\left. +\;2\Gamma _{\alpha \beta }^n\cdot D_n^{\left( 2q-2k\right) }\right]
_{\mu _2\ldots \mu _{2q+1-2k}}\;-\;\left( 2q-2k\right) \left( 2q-2k-1\right)
\cdot 
$$
$$
\cdot \left[ \left( emb^{*}D_\beta ^{\left( 2q-2k-2\right) }\right) _{\mu
_2\ldots \mu _{2q-1-2k}}\cdot \partial _{\mu _{2q-2k}}\left( emb^{*}\Delta
_\alpha \right) _{\mu _{2q+1-2k}}\right. \;+ 
$$
$$
\left. +\;\left. \left( emb^{*}D_\alpha ^{\left( 2q-2k-2\right) }\right)
_{\mu _2\ldots \mu _{2q-1-2k}}\cdot \partial _{\mu _{2q-2k}}\left(
emb^{*}\Delta _\beta \right) _{\mu _{2q+1-2k}}\right] \right\} \cdot \; 
$$
\begin{equation}
\label{an29}\cdot F_{\nu _1\nu _2}\cdots F_{\nu _{2k-1}\nu _{2k}}\quad . 
\end{equation}
\pagebreak[3]
The appropriate expression for $S_{\alpha m}$ is%
$$
S_{\alpha m}\left( U\right) \;=\;\frac{\epsilon ^{0\nu _1\ldots \nu _{2q}}}{%
2^q\cdot q!}\,emb^{*}\left[ \delta _\alpha D_m^{\left( 0\right) }+i\delta
_mD_\alpha ^{\left( 0\right) }\right] \cdot F_{\nu _1\nu _2}\cdots F_{\nu
_{2q-1}\nu _{2q}}\;+ 
$$
$$
+\;\sum\limits_{k=0}^{q-1}\frac{\epsilon ^{0\mu _2\ldots \mu _{2q+1-2k}\nu
_1\ldots \nu _{2k}}}{\left( 2q-2k\right) !\,2^k\cdot k!}\,\left\{
emb^{*}\left[ \delta _\alpha D_m^{\left( 2q-2k\right) }+i\delta _mD_\alpha
^{\left( 2q-2k\right) }\right] _{\mu _2\ldots \mu _{2q+1-2k}}\right. \;+ 
$$
$$
+\;\left( 2q-2k\right) \left( 2q-2k-1\right) \cdot 
$$
$$
\cdot \left[ \left( emb^{*}D_m^{\left( 2q-2k-2\right) }\right) _{\mu
_2\ldots \mu _{2q-1-2k}}\cdot \partial _{\mu _{2q-2k}}\left( emb^{*}\Delta
_\alpha \right) _{\mu _{2q+1-2k}}\right. \;+ 
$$
$$
\left. +\;i\cdot \left. \left( emb^{*}D_\alpha ^{\left( 2q-2k-2\right)
}\right) _{\mu _2\ldots \mu _{2q-1-2k}}\cdot \partial _{\mu _{2q-2k}}\left(
emb^{*}\Delta _m\right) _{\mu _{2q+1-2k}}\right] \right\} \cdot \; 
$$
\begin{equation}
\label{an30}\cdot F_{\nu _1\nu _2}\cdots F_{\nu _{2k-1}\nu _{2k}}\quad . 
\end{equation}
The expression for $S_{mn}$ is%
$$
S_{mn}\left( U\right) \;=\;-i\cdot \frac{\epsilon ^{0\nu _1\ldots \nu _{2q}}%
}{2^q\cdot q!}\,emb^{*}\left[ \delta _mD_n^{\left( 0\right) }-\delta
_nD_m^{\left( 0\right) }\right] \cdot F_{\nu _1\nu _2}\cdots F_{\nu
_{2q-1}\nu _{2q}}\;- 
$$
$$
-i\cdot \sum\limits_{k=0}^{q-1}\frac{\epsilon ^{0\mu _2\ldots \mu
_{2q+1-2k}\nu _1\ldots \nu _{2k}}}{\left( 2q-2k\right) !\,2^k\cdot k!}%
\,\left\{ emb^{*}\left[ \delta _mD_n^{\left( 2q-2k\right) }-\delta
_nD_m^{\left( 2q-2k\right) }\right] _{\mu _2\ldots \mu _{2q+1-2k}}\right.
\;+ 
$$
$$
\;+\;\left( 2q-2k\right) \left( 2q-2k-1\right) \cdot 
$$
$$
\cdot \left[ \left( emb^{*}D_n^{\left( 2q-2k-2\right) }\right) _{\mu
_2\ldots \mu _{2q-1-2k}}\cdot \partial _{\mu _{2q-2k}}\left( emb^{*}\Delta
_m\right) _{\mu _{2q+1-2k}}\right. \;- 
$$
$$
\left. -\;\left. \left( emb^{*}D_m^{\left( 2q-2k-2\right) }\right) _{\mu
_2\ldots \mu _{2q-1-2k}}\cdot \partial _{\mu _{2q-2k}}\left( emb^{*}\Delta
_n\right) _{\mu _{2q+1-2k}}\right] \right\} \cdot \; 
$$
\begin{equation}
\label{an31}\cdot F_{\nu _1\nu _2}\cdots F_{\nu _{2k-1}\nu _{2k}}\quad . 
\end{equation}

\subsection{D-$p$-branes in IIA supergravity backgrounds}

\subsubsection{Superspace constraints}

$D=10$ type II supergravity theories are the low-energy effective field
theories of type II superstring theories \cite{Thor}. These theories have
classical solutions which describe extended objects called $p$-branes. The $%
p $-branes are solitons carrying conserved charges that act as sources for
the various anti-symmetric gauge fields of the underlying supergravity
theory, i.e. RR-gauge fields $C^{\left( r\right) }$ and the NS-NS $2$-form
potential $B$.

In superspace all ordinary components of RR and NS-NS gauge fields are
introduced as first components of their corresponding superfields. From the
gauge fields one can derive field strengths with associated Bianchi
identities. In order to reduce the enormous field content of these
superfields down to the on-shell content one introduces {\it constraints} on
some of the components of the superfield field strengths. When these
constraints are inserted into the Bianchi identities the latter cease to be
identities, but rather become equations the consistency of which has to be
examined separately. If the constraints are properly chosen the equations so
obtained are just the supergravity equations of motion.

D-branes arise from prescribing mixed (Neumann- and Dirichlet) boundary
conditions on open strings in type II string theory. They are introduced as $%
\left( p+1\right) $-dimensional hypersurfaces in spacetime where open
strings are constrained to end on, but the ends are free to move on this
submanifold. A spacelike section of a D-brane can be given a finite volume
in a spacetime with compact dimensions by wrapping around topologically
non-trivial cycles in the spacetime. In this case the supertranslation
algebra of Noether charges or modified charges carried by the brane is
extended by topological charges, which we derive below.

We want to consider D-branes in a flat IIA background. This condition
requires the underlying supergravity theory to be massless, $m=0$, since it
is known that $D=10$ Minkowski spacetime is {\bf not} a solution to the
field equations of massive IIA supergravity \cite{Berg2}. The massless
theory allows a flat solution, however; its constraints, i.e. the
constraints on the massless IIA supergravity background, can be obtained by
dimensional reduction of the standard $D=11$ superspace constraints \cite
{Berg3}; in particular, they imply the field equations of massless IIA
supergravity. Moreover, we have the observation that, once the constraints
on the NS-NS fields coupling to the kinetic (supersymmetry invariant) term
in the D-brane action are given, the constraints on the RR-fields coupling
to the brane via the Wess-Zumino term can be read off from $\kappa $%
-symmetry, see \cite{Ceder1}; thus, consistent propagation of $D$-branes
demands a background solving the equations of motion of the appropriate
supergravity theory.

\subsubsection{Superspace background and Bianchi identities \label{Vakuum}}

In the following we choose a massless flat background vacuum with Dilaton $%
\phi =0$, Dilatino $D\phi =0$, where $D$ denotes a supercovariant
derivative. Moreover, we assume that all bosonic components of the field
strengths associated with the NS-NS fields and RR gauge fields,
respectively, are zero; the non-bosonic components of these field strengths
as well as the non-bosonic torsion components are uniquely determined by the
superspace constraints, see \cite{BergTown1}. The RR superfield potentials
are usually collected in a formal sum $C=\sum_{r=0}^{10}C^{\left( r\right) }$%
, where the ordinary RR gauge potentials are just the leading components of
the $C^{\left( r\right) }$ in a $\theta $-expansion; for the IIA case only
the odd forms are relevant. Their field strengths are defined in (\ref{lag2}%
). The Bianchi identities associated with these field strengths are given in
(\ref{lag3}). The Bianchi identities for the field strength $H$ of the NS-NS
field $B$ is $dH=0$.

Let us now define a family of superspace forms $K^{\left( p+2\right) }\left(
S\right) $ by 
\begin{equation}
\label{ap13}K^{\left( p+2\right) }\left( S\right) =\frac i{p!}\Pi
^{m_p}\cdots \Pi ^{m_1}\cdot d\bar \theta S\Gamma _{m_1\ldots m_p}d\theta
\quad , 
\end{equation}
where $p=0,\ldots ,9$, $S\in \left\{ {\bf 1}_{32},\Gamma _{11}\right\} $,
and $\Gamma _{m_1\ldots m_p}$ is the usual antisymmetrised product of $%
\Gamma $-matrices. Then our choice of vacuum determines the field strengths
to be \cite{BergTown1} 
\begin{equation}
\label{ap14}
\begin{array}{c}
R^{\left( 2\right) }=K^{\left( 2\right) }\left( \Gamma _{11}\right) \quad ,
\\ 
R^{\left( 4\right) }=K^{\left( 4\right) }\left( 
{\bf 1}\right) \quad , \\ R^{\left( 6\right) }=K^{\left( 6\right) }\left(
\Gamma _{11}\right) \quad , \\ 
R^{\left( 8\right) }=K^{\left( 8\right) }\left( 
{\bf 1}\right) \quad , \\ R^{\left( 10\right) }=K^{\left( 10\right) }\left(
\Gamma _{11}\right) \quad ; 
\end{array}
\end{equation}
furthermore, the field strength $H$ must take the form 
\begin{equation}
\label{ap15}H=-K^{\left( 3\right) }\left( \Gamma _{11}\right) \quad . 
\end{equation}
These field strengths are determined by superspace constraints; the Bianchi
identities (\ref{lag3}) and the relation $dH=0$ are therefore identities no
longer, and we must check whether they are actually satisfied.

\subsubsection{Explicit form of $B$}

Let us first consider the field strength $H$ in (\ref{ap15}); the $3$-form $%
K^{\left( 3\right) }\left( \Gamma _{11}\right) $ has a potential 
\begin{equation}
\label{ap151}B=\left( -\Pi ^m+\frac i2d\bar \theta \Gamma ^m\theta \right)
\cdot \left( id\bar \theta \Gamma _{11}\Gamma _m\theta \right) \quad , 
\end{equation}
which yields $H=dB=-K^{\left( 3\right) }\left( \Gamma _{11}\right) $ on
account of the identity 
\begin{equation}
\label{ap152}d\bar \theta \Gamma ^n\theta \cdot d\bar \theta \Gamma
_{11}\Gamma _nd\theta +d\bar \theta \Gamma _{11}\Gamma _n\theta \cdot d\bar
\theta \Gamma ^nd\theta =0\quad ; 
\end{equation}
this in turn is a consequence of the identity 
\begin{equation}
\label{ap153}\Gamma _{(\alpha \beta }^n\left( \Gamma _{11}\Gamma _n\right)
_{\gamma \delta )}=0\quad , 
\end{equation}
which is known to to hold in $D=10$. Therefore (\ref{ap15}) actually is a
consistent choice for $H$. Since $H=-K^{\left( 3\right) }\left( \Gamma
_{11}\right) $ is indeed supertranslation invariant we have $\delta _\alpha
B=d\Delta _\alpha $ for some $1$-form $\Delta _\alpha $; this can be
computed to be 
\begin{equation}
\label{ap154}\Delta _\alpha =dX^m\cdot \left( i\Gamma _{11}\Gamma _m\theta
\right) _\alpha -\frac 16\left[ d\bar \theta \Gamma _m\theta \cdot \left(
\bar \theta \Gamma _{11}\Gamma _m\right) _\alpha +d\bar \theta \Gamma
_{11}\Gamma _m\theta \cdot \left( \bar \theta \Gamma _m\right) _\alpha
\right] \quad . 
\end{equation}
Moreover we note that%
$$
\delta _\alpha \Delta _\beta +\delta _\beta \Delta _\alpha =dX^m\cdot \left(
2i\Gamma _{11}\Gamma _m\right) _{\alpha \beta }\;+ 
$$
\begin{equation}
\label{ap155}+\;\frac 12\cdot d\left[ \left( \Gamma _{11}\Gamma _m\theta
\right) _\alpha \cdot \left( \Gamma ^m\theta \right) _\beta +\left( \Gamma
_{11}\Gamma _m\theta \right) _\beta \cdot \left( \Gamma ^m\theta \right)
_\alpha \right] \quad . 
\end{equation}
Taking the trace with $\Gamma _n^{\alpha \beta }$ of this expression yields
zero: due to the tracelessness of products of $\Gamma $-matrices the first
contribution vanishes, and the terms in the square bracket yield zero since 
$$
\bar \theta \Gamma _{11}\Gamma _m\Gamma _n\Gamma ^m\theta =\left( 2-D\right)
\bar \theta \Gamma _{11}\Gamma _n\theta =0\quad , 
$$
for $C\Gamma _{11}\Gamma _n$ is {\bf symmetric}, see Table \ref{Tabelle2}.
By (\ref{an4}) this implies that 
\begin{equation}
\label{ap156}\Delta _m=0\quad , 
\end{equation}
as can be seen directly from (\ref{ap151}), since $B$ is translation
invariant. From definitions (\ref{an13}, \ref{an19}, \ref{an21}) we now see
that 
\begin{equation}
\label{for1}S_{\alpha \beta }\left( \Delta \right) =-dX^m\cdot \left(
2i\Gamma _{11}\Gamma _m\right) _{\alpha \beta }\;+\;\cdots \quad , 
\end{equation}
\begin{equation}
\label{for2}S_{\alpha m}\left( \Delta \right) =S_{mn}\left( \Delta \right)
=0\quad , 
\end{equation}
where $"\cdots "$ denotes terms that involve only fermionic $1$-forms.
Finally, note that $d\Delta _m=0$.

\subsubsection{Generalized $\Gamma $-matrix identities}

Now let us turn attention to the RR field strengths. If we insert (\ref{ap14}%
) into the Bianchi identities (\ref{lag3}) we obtain 
\begin{equation}
\label{ap16}dK^{\left( 2\right) }\left( \Gamma _{11}\right) =0\quad ,
\end{equation}
\begin{equation}
\label{ap17}
\begin{array}{c}
dK^{\left( 4\right) }\left( 
{\bf 1}\right) +K^{\left( 2\right) }\left( \Gamma _{11}\right) K^{\left(
3\right) }\left( \Gamma _{11}\right) =0\quad , \\ dK^{\left( 6\right)
}\left( \Gamma _{11}\right) +K^{\left( 4\right) }\left( 
{\bf 1}\right) K^{\left( 3\right) }\left( \Gamma _{11}\right) =0\quad , \\ 
dK^{\left( 8\right) }\left( 
{\bf 1}\right) +K^{\left( 6\right) }\left( \Gamma _{11}\right) K^{\left(
3\right) }\left( \Gamma _{11}\right) =0\quad , \\ dK^{\left( 10\right)
}\left( \Gamma _{11}\right) +K^{\left( 8\right) }\left( {\bf 1}\right)
K^{\left( 3\right) }\left( \Gamma _{11}\right) =0\quad .
\end{array}
\end{equation}
Here (\ref{ap16}) is trivially satisfied due to $d\left( id\bar \theta
\Gamma _{11}d\theta \right) =0$. The equations in (\ref{ap17}) can be
written as 
\begin{equation}
\label{ap18}dK^{\left( 2q+2\right) }\left( S\right) +K^{\left( 2q\right)
}\left( S\Gamma _{11}\right) K^{\left( 3\right) }\left( \Gamma _{11}\right)
=0\quad ;\quad q=1,2,3,4,
\end{equation}
and $p=2q$ is related to $S$ by Table \ref{Tabelle1}. {%
\begin{table} \centering
   \begin{tabular}{|c||c|c|c|c|c|} \hline
   ${\bf {p=2q}}$  &  $0$  &  $2$  &  $4$  &  $6$  &  $8$  \\ \hline
   ${\bf S}$  &  $\Gamma _{11}$  &  ${\bf 1}_{32}$  & $\Gamma _{11}$  & ${\bf 1}_{32}$  &  $\Gamma _{11}$    \\ \hline
  \end{tabular}
 \caption{Relation between $p$ and $S$.  \label{Tabelle1}}
\end{table}} If we now use the explicit definitions of $K^{\left( 2q\right)
}\left( S\right) $ as given in (\ref{ap13}) we find that equations (\ref
{ap17}) are satisfied {\bf iff} 
\begin{equation}
\label{ap19}\Gamma _{(\alpha \beta }^n\left( S\Gamma _{nm_1\ldots
m_{2q-1}}\right) _{\gamma \delta )}\;+\;\left( 2q-1\right) \cdot \left(
\Gamma _{11}\Gamma _{[m_1}\right) _{(\alpha \beta }\left( S\Gamma
_{11}\Gamma _{m_2\ldots m_{2q-1}]}\right) _{\gamma \delta )}\;=\;0\quad .
\end{equation}
In the first term the symmetrisation involves spinor indices $\alpha ,\beta
,\gamma ,\delta $, but {\bf no} covector indices $n,m_1,\ldots ,m_{2q-1}$,
of course. In the second term we have a symmetrisation over $\alpha ,\beta
,\gamma ,\delta $, and independently, an antisymmetristion over $m_1,\ldots
m_{2q-1}$. (\ref{ap19}) is a set of {\it generalized }$\Gamma ${\it -matrix
identities}. We shall derive a necessary condition for them to hold, and
show, that it is indeed satisfied. Before we do so, however, let us examine
the special case of (\ref{ap19}) when $q=1$. In this case (\ref{ap19})
becomes (see table \ref{Tabelle1} for the choice of $S$) 
\begin{equation}
\label{ap20}\Gamma _{(\alpha \beta }^n\left( \Gamma _{nm}\right) _{\gamma
\delta )}+\left( \Gamma _{11}\right) _{(\alpha \beta }\left( \Gamma
_{11}\Gamma _m\right) _{\gamma \delta )}\;=\;0\quad .
\end{equation}
This is just the dimensional reduction to $D=10$ of the $D=11$ identity
required for $\kappa $-symmetry of the $D=11$ supermembrane \cite{Berg3},
and is known to hold in $D=11$. This means that the validity of at least the
first equation in (\ref{ap17}) is assured.

To examine the validity of the other cases we reexpress (\ref{ap19}) as%
$$
\Gamma _{(\alpha \beta }^n\left( S\Gamma _{nm_1\ldots m_{2q-1}}\right)
_{\gamma \delta )}+\left( \Gamma _{11}\Gamma _{m_1}\right) _{(\alpha \beta
}\left( S\Gamma _{11}\Gamma _{m_2\ldots m_{2q-1}}\right) _{\gamma \delta )}+ 
$$
\begin{equation}
\label{ap21}+\;\left( \mbox{cycl. }m_1\rightarrow m_2\rightarrow \cdots
\right) \;+\;\cdots \;=0\quad , 
\end{equation}
where ''cyc.'' denotes a sum over all cyclic permutations of $m_i$-indices
in the second term of the first line. Now we multiply (\ref{ap21}) by $%
\Gamma _{\alpha \beta }^l$; this yields%
$$
tr\left( \Gamma ^n\Gamma _l\right) \cdot \left( CS\Gamma _{nm_1\ldots
m_{2q-1}}\right) +\left( C\Gamma ^n\right) \cdot tr\left( \Gamma _lS\Gamma
_{nm_1\ldots m_{2q-1}}\right) + 
$$
$$
+4\left( C\Gamma ^n\Gamma _lS\Gamma _{nm_1\ldots m_{2q-1}}\right) _{\left(
sym\right) }+ 
$$
$$
+\left\{ tr\left( \Gamma _{11}\Gamma _{m_1}\Gamma _l\right) \cdot \left(
CS\Gamma _{11}\Gamma _{m_2\ldots m_{2q-1}}\right) +\left( C\Gamma
_{11}\Gamma _{m_1}\right) \cdot tr\left( \Gamma _lS\Gamma _{11}\Gamma
_{m_2\ldots m_{2q-1}}\right) +\right. 
$$
\begin{equation}
\label{ap22}+\left. 4\left( C\Gamma _{11}\Gamma _{m_1}\Gamma _lS\Gamma
_{11}\Gamma _{m_2\ldots m_{2q-1}}\right) _{\left( sym\right) }+\left( \mbox{%
cycl. }m_1\rightarrow m_2\rightarrow \cdots \right) \;+\;\cdots \right\}
=0\quad . 
\end{equation}
Here $\left( sym\right) $ denotes the symmetric part of the matrix in
brackets, i.e. $M_{\left( sym\right) }=\frac{1}{2}\left( M+M^T\right) 
$. We list the contributions to (\ref{ap22}):%
$$
tr\left( \Gamma ^n\Gamma _l\right) \cdot \left( CS\Gamma _{nm_1\ldots
m_{2q-1}}\right) =tr\left( {\bf 1}_{32}\right) \cdot \left( CS\Gamma
_{lm_1\ldots m_{2q-1}}\right) \quad , 
$$
$$
tr\left( \Gamma _lS\Gamma _{nm_1\ldots m_{2q-1}}\right) =0\quad \mbox{for
all\quad }q=1,\ldots ,4\,;\;S=S\left( q\right) \mbox{, see Table (\ref
{Tabelle1})\quad }, 
$$
\begin{equation}
\label{ap23}\left( C\Gamma ^n\Gamma _lS\Gamma _{nm_1\ldots m_{2q-1}}\right)
_{\left( sym\right) }=-\left( D-2q-1\right) \cdot \left( CS\Gamma
_{lm_1\ldots m_{2q-1}}\right) \quad , 
\end{equation}
where we have used the fact that if $\left( CS\Gamma _{lm_1\ldots
m_{2q-1}}\right) $ is symmetric then $\left( CS\Gamma _{m_2\ldots
m_{2q-1}}\right) $ is always antisymmetric, see Table \ref{Tabelle2}.{%
\begin{table} \centering
  \begin{tabular}{||c||c|c||c|c||} \hline
     ${\bf p}$  &  $S={\bf 1}_{32}$  &  type  &  $S=\Gamma _{11}$  &  type  \\ \hline \hline
     $0$  &  $C$  &  $-$  &  $C\Gamma _{11}$  &  $+$  \\  \hline 
     $1$  &  $C\Gamma _{m_1}$  &  $+$  &  $C\Gamma _{11}\Gamma _{m_1}$
  &  $+$  \\  \hline
     $2$  &  $C\Gamma _{m_1m_2}$  &  $+$  &  $C\Gamma _{11}\Gamma _{m_1m_2}$  &  $-$  \\  \hline
     $3$  &  $C\Gamma _{m_{1\ldots }m_3}$  &  $-$  &  $C\Gamma _{11}\Gamma _{m_{1\ldots }m_3}$  &  $-$  \\  \hline
     $4$  &  $C\Gamma _{m_{1\ldots }m_4}$  &  $-$  &  $C\Gamma _{11}\Gamma _{m_{1\ldots }m_4}$  &  $+$  \\  \hline
     $5$  &  $C\Gamma _{m_{1\ldots }m_5}$  &  $+$  &  $C\Gamma _{11}\Gamma _{m_{1\ldots }m_5}$  &  $+$  \\  \hline
     $6$  &  $C\Gamma _{m_{1\ldots }m_6}$  &  $+$  &  $C\Gamma _{11}\Gamma _{m_{1\ldots }m_6}$  &  $-$  \\  \hline
     $7$  &  $C\Gamma _{m_{1\ldots }m_7}$  &  $-$  &  $C\Gamma _{11}\Gamma _{m_{1\ldots }m_7}$  &  $-$  \\  \hline
     $8$  &  $C\Gamma _{m_{1\ldots }m_8}$  &  $-$  &  $C\Gamma _{11}\Gamma _{m_{1\ldots }m_8}$  &  $+$  \\  \hline
     $9$  &  $C\Gamma _{m_{1\ldots }m_9}$  &  $+$  &  $C\Gamma _{11}\Gamma _{m_{1\ldots }m_9}$  &  $+$  \\  \hline
     $10$  &  $C\Gamma _{m_{1\ldots }m_{10}}$  &  $+$  &  $C\Gamma _{11}\Gamma _{m_{1\ldots }m_{10}}$  &  $-$  \\  \hline
  \end{tabular}
  \caption{Symmetry and Antisymmetry of products of $\Gamma$-matrices in $D=10$. $+/-$ denotes Symmetry/Antisymmetry; $C$ is a charge conjugation matrix.\label{Tabelle2}}
\end{table}} The last three contributions to (\ref{ap22}) are%
$$
tr\left( \Gamma _{11}\Gamma _{m_1}\Gamma _l\right) =0\quad , 
$$
$$
tr\left( \Gamma _lS\Gamma _{11}\Gamma _{m_2\ldots m_{2q-1}}\right) =0\quad , 
$$
$$
\left( CS\Gamma _{m_1}\Gamma _l\Gamma _{m_2\ldots m_{2q-1}}\right) _{\left(
sym\right) }\;=\;-\left( CS\Gamma _{lm_1\ldots m_{2q-1}}\right) - 
$$
\begin{equation}
\label{ap24}-\left( 2q-1\right) \left( 2q-3\right) \cdot \eta
_{m_1[m_2}\cdot \eta _{\left| l\right| m_3}\left( CS\Gamma _{m_4\ldots
m_{2q-1}]}\right) \quad . 
\end{equation}
Now we must perform the cyclic sum $\left( \mbox{cycl. }m_1\rightarrow
m_2\rightarrow \cdots \right) $ in (\ref{ap24}). Since this is equal to $%
\left( 2q-1\right) \times $ ''antisymmetrisation of (\ref{ap24}) over $%
\left( m_1,\ldots ,m_{2q-1}\right) $'' we see that the second contribution
on the right hand side of (\ref{ap24}) must vanish, since it involves
antisymmetrisation over $\eta _{m_1m_2}$, and therefore the total
contribution from this term is
\begin{equation}
\label{ap25}\left( 2q-1\right) \cdot \left( CS\Gamma _{[m_1}\Gamma _{\left|
l\right| }\Gamma _{m_2\ldots m_{2q-1}]}\right) _{\left( sym\right) }=-\left(
2q-1\right) \cdot \left( CS\Gamma _{lm_1\ldots m_{2q-1}}\right) \quad . 
\end{equation}
Altogether, (\ref{ap22}) leads to the condition 
\begin{equation}
\label{ap251}\left[ tr\left( {\bf 1}_{32}\right) -4\left( D-2q-1\right)
-4\left( 2q-1\right) \right] \cdot \left( CS\Gamma _{lm_1\ldots
m_{2q-1}}\right) =0\quad ; 
\end{equation}
remarkably, the contributions involving $q$ cancel each other in this
equation, so we arrive at 
\begin{equation}
\label{ap26}tr\left( {\bf 1}_{32}\right) -4\left( D-2\right) =0 
\end{equation}
as a necessary condition for the $\Gamma $-matrix identities (\ref{ap19}) to
hold; but this is satisfied precisely in $D=10$, {\bf independent} of $q$.

We do not know whether (\ref{ap26}) is also sufficient to ensure (\ref{ap19}%
); in the past, sufficiency of a similar condition to (\ref{ap26}) to
establish the well-known $\Gamma $-matrix identity $\Gamma _{\left( \alpha
\beta \right) }^n\left( \Gamma _n\right) _{\gamma \delta )}=0$ in $D=10$
could be established only via computer \cite{PKT1}. In the following we
shall assume that (\ref{ap26}) is sufficient and therefore (\ref{ap19})
holds for all allowed values of $q$; if this assumption should turn out to
be wrong, then at least our analysis is valid for $q=1$, since in this case
the validity of (\ref{ap20}) is known; our results then would be restricted
to the D-$2$-brane in a IIA superspace.

\subsubsection{Constructing the leading terms of $C^{\left( r\right) }$}

Provided that (\ref{ap19}) is valid we show that under these circumstances
we can construct the potentials $C^{\left( 3\right) },C^{\left( 5\right)
},C^{\left( 7\right) },C^{\left( 9\right) }$ recursively from $C^{\left(
1\right) }$. From (\ref{ap13}, \ref{ap14}) we see that, up to a gauge
transformation, we have 
\begin{equation}
\label{ap27}C^{\left( 1\right) }=id\bar \theta \Gamma _{11}\theta \quad . 
\end{equation}
Now assuming that we have constructed $C^{\left( 2q-1\right) }$ we can use (%
\ref{lag2}) to give 
\begin{equation}
\label{ap28}dC^{\left( 2q+1\right) }=K^{\left( 2q+2\right) }\left( S\right)
-C^{\left( 2q-1\right) }K^{\left( 3\right) }\left( \Gamma _{11}\right) \quad
, 
\end{equation}
where $S$ is chosen according to Table \ref{Tabelle1}. A nessecary and
sufficient condition for the existence of a (local) $\left( 2q+1\right) $%
-form $C^{\left( 2q+1\right) }$ that satisfies (\ref{ap28}) is that the
differential of the right hand side of (\ref{ap28}) vanishes; but since $%
dC^{\left( 2q-1\right) }=K^{\left( 2q\right) }\left( S\Gamma _{11}\right)
-C^{\left( 2q-3\right) }K^{\left( 3\right) }\left( \Gamma _{11}\right) $ by
assumption, this is 
\begin{equation}
\label{ap29}d\left[ K^{\left( 2q+2\right) }\left( S\right) -C^{\left(
2q-1\right) }K^{\left( 3\right) }\left( \Gamma _{11}\right) \right]
=dK^{\left( 2q+2\right) }\left( S\right) +K^{\left( 2q\right) }\left(
S\Gamma _{11}\right) K^{\left( 3\right) }\left( \Gamma _{11}\right) \quad , 
\end{equation}
where we have used the fact that $H=-K^{\left( 3\right) }\left( \Gamma
_{11}\right) $ is a closed $3$-form. But the right hand side of (\ref{ap29})
are just the Bianchi identities (\ref{ap18}), which are identically zero
provided that (\ref{ap19}) holds; the Bianchi identities are therefore
integrability conditions for the forms $C^{\left( 2q+1\right) }$ in (\ref
{ap28}). The existence of $C^{\left( r\right) }$ is therefore guaranteed at
least for $r=1,3$.

We have solved (\ref{ap28}) for $C^{\left( 3\right) }$ explicitly; the
result is%
$$
C^{\left( 3\right) }=\frac i2\Pi ^m\Pi ^n\cdot d\bar \theta \Gamma
_{nm}\theta \;+ 
$$
$$
+\;\frac 12\Pi ^m\cdot \left[ d\bar \theta \Gamma _n\theta \cdot d\bar
\theta \Gamma _{nm}\theta -d\bar \theta \Gamma _{11}\theta \cdot d\bar
\theta \Gamma _{11}\Gamma _m\theta \right] \;+ 
$$
\begin{equation}
\label{ap30}+\;\frac i6d\bar \theta \Gamma _m\theta \cdot \left[ d\bar
\theta \Gamma _{11}\theta \cdot d\bar \theta \Gamma _{11}\Gamma _m\theta
-d\bar \theta \Gamma ^n\theta \cdot d\bar \theta \Gamma _{nm}\theta \right]
\quad . 
\end{equation}
In proving that (\ref{ap30}) is actually a solution to (\ref{ap28}) for $q=1$
one has to make use of the identities 
\begin{equation}
\label{ap31}\left( Id\,1\right) :=d\bar \theta \Gamma ^nd\theta \cdot d\bar
\theta \Gamma _{11}\Gamma _n\theta +d\bar \theta \Gamma ^n\theta \cdot d\bar
\theta \Gamma _{11}\Gamma _nd\theta =0\quad , 
\end{equation}
and%
$$
\left( Id\,2\right) _m:=d\bar \theta \Gamma ^nd\theta \cdot d\bar \theta
\Gamma _{nm}\theta +d\bar \theta \Gamma ^n\theta \cdot d\bar \theta \Gamma
_{nm}d\theta \;+ 
$$
\begin{equation}
\label{ap32}+\;d\bar \theta \Gamma _{11}d\theta \cdot d\bar \theta \Gamma
_{11}\Gamma _m\theta +d\bar \theta \Gamma _{11}\theta \cdot d\bar \theta
\Gamma _{11}\Gamma _md\theta \;=\;0\quad , 
\end{equation}
where (\ref{ap31}) is a consequence of (\ref{ap153}), and (\ref{ap32})
follows from (\ref{ap20}). Then%
$$
dC^{\left( 3\right) }=K^{\left( 4\right) }\left( {\bf 1}_{32}\right)
-C^{\left( 1\right) }K^{\left( 3\right) }\left( \Gamma _{11}\right) \;+ 
$$
$$
+\;\left( \frac 12\Pi ^m-\frac i6d\bar \theta \Gamma ^m\theta \right) \cdot
\left( Id\,2\right) _m\;-\;\left( \frac i3d\bar \theta \Gamma _{11}\theta
\right) \cdot \left( Id\,1\right) \quad , 
$$
and (\ref{ap28}) is fulfilled.

In principle we could apply the same procedure to construct the other
potentials $C^{\left( 5\right) },C^{\left( 7\right) },C^{\left( 9\right) }$.
But for the purpose we are pursuing here, namely the determination of the
topological extensions of Noether algebras, we do not need to know the full
expression for $C^{\left( 2q+1\right) }$; as mentioned earlier, these
algebra extensions come into play when the D-$p$-brane wraps around compact
dimensions in the spacetime; but the topology of this configuration is
entirely determined by the bosonic coordinates $X$ on the superspace, and
the pull-back of the differentials $dX^m$ to the worldvolume of the brane,
respectively. In evaluating the anomalous contributions to the charge
algebra as far as they origin in the WZ-term we therefore can restrict
attention to those components of the $C$'s which have only bosonic indices.
The strategy is as follows:

From section \ref{Modificatio} we see that all we need are the components of
the forms $S_{MN}\left( U\right) $, $M=\left( m,\alpha \right) $, carrying
the maximum number of bosonic indices; since we shall work with the LI-basis
now, this means that we need only consider terms involving the maximum
number of bosonic basis-$1$-forms $\Pi ^m$; in the following we shall refer
to such terms simply as ''leading terms''; furthermore we shall call the
number of bosonic indices in the leading term as the ''order'' of the term.
From (\ref{an29})-(\ref{an31}) we see that $S_{MN}\left( U\right) $ is
composed of terms $\delta _MD_N$, $D_M$ and $D_Md\Delta _N$. Since $\delta
_M $ leaves the number of LI-$1$-forms invariant we see that in order to
construct the leading terms of $S_{MN}\left( U\right) $ we need only
construct the leading terms of $D_M$. Now let us look back at formula (\ref
{lag7}), 
\begin{equation}
\label{bos2}\delta _\alpha C^{\left( 2q+1\right)
}=\sum\limits_{k=0}^qdD_\alpha ^{\left( 2q-2k\right) }\cdot \frac{B^k}{k!}%
\quad . 
\end{equation}
From our choice of $B$ in (\ref{ap151}) we see that $B$ contains only one
bosonic $1$-form $\Pi ^m$; the order of the terms in the sum in (\ref{bos2})
therefore decreases by $1$ as $k$ increases by $1$; this means that in order
to construct the leading term of $dD_\alpha ^{\left( 2q\right) }$ we need
only construct the leading term in $\delta _\alpha C^{\left( 2q+1\right) }$;
but this can be done using (\ref{ap28}) recursively: 
\begin{equation}
\label{bos2x}dC^{\left( 2q+1\right) }=K^{\left( 2q+2\right) }\left( S\right)
-C^{\left( 2q-1\right) }K^{\left( 3\right) }\left( \Gamma _{11}\right) \quad
. 
\end{equation}
From (\ref{ap13}) we see that the order of $K^{\left( 3\right) }\left(
\Gamma _{11}\right) $ is one, and that of $K^{\left( 2q+2\right) }\left(
S\right) $ is $\left( 2q\right) $; from (\ref{ap27}) and (\ref{ap30}) we
deduce that the order of $C^{\left( 2q-1\right) }$ is $\left( 2q-2\right) $,
therefore the first term on the right hand side of (\ref{bos2x}) is the
leading term, and we must construct a $C^{\left( 2q+1\right) }$ such that%
$$
dC^{\left( 2q+1\right) }=K^{\left( 2q+2\right) }\left( S\right) +\cdots
\quad . 
$$
Thus we find the leading term of $C^{\left( 2q+1\right) }$ to be 
\begin{equation}
\label{bos3}C^{\left( 2q+1\right) }=\frac i{\left( 2q\right) !}\Pi
^{m_{2q}}\cdots \Pi ^{m_1}\cdot d\bar \theta S\Gamma _{m_1\ldots
m_{2q}}\theta \quad , 
\end{equation}
with $S$ given in Table \ref{Tabelle1}. Therefore the leading term of $%
dD_\alpha ^{\left( 2q\right) }$ is 
\begin{equation}
\label{bos4}dD_\alpha ^{\left( 2q\right) }=-\frac i{\left( 2q\right) !}\Pi
^{m_{2q}}\cdots \Pi ^{m_1}\cdot \left( d\bar \theta S\Gamma _{m_1\ldots
m_{2q}}\right) _\alpha \quad , 
\end{equation}
and, up to a differential, we have 
\begin{equation}
\label{bos5}D_\alpha ^{\left( 2q\right) }=-\frac i{\left( 2q\right) !}\Pi
^{m_{2q}}\cdots \Pi ^{m_1}\cdot \left( \bar \theta S\Gamma _{m_1\ldots
m_{2q}}\right) _\alpha \quad . 
\end{equation}
This gives 
\begin{equation}
\label{bos6}\delta _\alpha D_\beta ^{\left( 2q\right) }+\delta _\beta
D_\alpha ^{\left( 2q\right) }=\frac{-2i}{\left( 2q\right) !}\Pi
^{m_{2q}}\cdots \Pi ^{m_1}\cdot \left( S\Gamma _{m_1\ldots m_{2q}}\right)
_{\alpha \beta }\quad = 
\end{equation}
$$
=\;\frac{-2i}{\left( 2q\right) !}dX^{m_1}\cdots dX^{m_{2q}}\cdot \left(
S\left( 2q\right) \Gamma _{m_{2q}\ldots m_1}\right) _{\alpha \beta
}\;+\;\cdots \quad . 
$$
multiplying (\ref{bos6}) with $\Gamma _n^{\alpha \beta }$ then yields a
vanishing result due to the vanishing of the trace%
$$
tr\left( S\Gamma _{m_1\ldots m_{2q}}\Gamma _n\right) =0 
$$
for all allowed values of $q$, $n$ and $S$. But since the order of the
leading term of $D_\alpha ^{\left( 2q-2\right) }\cdot d\Delta _\beta $ is $%
\left( 2q-1\right) $, and the order of $\delta _\alpha D_\beta ^{\left(
2q\right) }$ is $\left( 2q\right) $, as can be seen from (\ref{ap154}) and (%
\ref{bos5}), we infer from (\ref{lag10}) that indeed 
\begin{equation}
\label{bos7}D_m^{\left( 0\right) }=0\quad ,\quad D_m^{\left( 2q\right)
}=0\quad ; 
\end{equation}
these equations will actually hold in a rigorous sense, not only to leading
order; from (\ref{ap27}) and (\ref{ap30}) we see that at least $C^{\left(
1\right) }$ and $C^{\left( 3\right) }$ are strictly translation invariant,
and this will be true for the others as well, since the higher rank
potentials are constructed recursively from the lower rank ones.
Furthermore, from (\ref{bos5}) we infer that $\delta _mD_\alpha ^{\left(
2q\right) }=0$ for all $q$.

\subsubsection{Extended superalgebras for D-$2q$-branes}

Now we can turn to evaluating the expressions $S_{MN}\left( U\right) $ as
given in (\ref{an29})-(\ref{an31}). Since expressions involving $d\Delta _M$
have leading order smaller than the leading order of $\delta _\alpha D_\beta
^{\left( 2q-2k\right) }$, see (\ref{for1}, \ref{for2}), they can be omitted
in the discussion. The final expression for $S_{\alpha \beta }\left(
U\right) $ is therefore%
$$
S_{\alpha \beta }\left( U\right) \;=\;-2i\cdot \sum\limits_{k=0}^q\left[
S\left( 2q-2k\right) \Gamma _{m_{2q-2k}\ldots m_1}\right] _{\alpha \beta
}\cdot \frac{\epsilon ^{0\mu _1\ldots \mu _{2q-2k}\nu _1\ldots \nu _{2k}}}{%
\left( \left( 2q-2k\right) !\right) ^2\,2^k\cdot k!}\,\cdot 
$$
\begin{equation}
\label{bos8}\cdot \partial _{\mu _1}X^{m_1}\cdots \partial _{\mu
_{2q-2k}}X^{m_{2q-2k}}\cdot F_{\nu _1\nu _2}\cdots F_{\nu _{2k-1}\nu
_{2k}}\quad . 
\end{equation}
Let us now write $dX^m:=emb^{*}dX^m$ for the sake of convenience; then we
have%
$$
dX^{m_1}\cdots dX^{m2q-2k}\cdot \frac{\left( dA\right) ^k}{k!}= 
$$
$$
=\;\omega _0\cdot \frac{\epsilon ^{0\mu _1\ldots \mu _{2q-2k}\nu _1\ldots
\nu _{2k}}}{\left( 2q-2k\right) !\,2^k\cdot k!}\cdot \partial _{\mu
_1}X^{m_1}\cdots \partial _{\mu _{2q-2k}}X^{m_{2q-2k}}\cdot F_{\nu _1\nu
_2}\cdots F_{\nu _{2k-1}\nu _{2k}}\quad , 
$$
where $\omega _0=d\sigma ^1\cdots d\sigma ^{2q}$; therefore 
\begin{equation}
\label{bos9}\omega _0\cdot S_{\alpha \beta }\left( U\right) =-2i\cdot
\sum\limits_{k=0}^q\frac{\left[ S\left( 2q-2k\right) \Gamma
_{m_{2q-2k}\ldots m_1}\right] _{\alpha \beta }}{\left( 2q-2k\right) !}\cdot
dX^{m_1}\cdots dX^{m2q-2k}\cdot \frac{\left( dA\right) ^k}{k!}\quad . 
\end{equation}

Furthermore, from (\ref{an30}) and (\ref{an31}) we infer that both $%
S_{\alpha m}\left( U\right) $ and $S_{mn}\left( U\right) $ are zero.

Now we can collect everything together to write down the general structure
of the modified charge algebra; we assume that double integration is
defined, so that we get%
$$
\left\{ Q_\alpha ,Q_\beta \right\} =2\Gamma _{\alpha \beta }^n\cdot
P_m\;-\;2i\left( \Gamma _{11}\Gamma _m\right) _{\alpha \beta }\cdot Y^m\;- 
$$
\begin{equation}
\label{bos10}-\;2i\cdot \sum\limits_{k=0}^q\frac{\left[ S\left( 2q-2k\right)
\Gamma _{m_{2q-2k}\ldots m_1}\right] _{\alpha \beta }}{\left( 2q-2k\right) !}%
\cdot Z^{m_1\ldots m_{2q-2k}}\quad , 
\end{equation}
with 
\begin{equation}
\label{bos11}Y^m=\frac 1{\left( 2q-1\right) !}\int\limits_{W\left( t\right)
}\left( *\Lambda ^{gauge}\right) \,dX^m\quad , 
\end{equation}
which was defined in (\ref{an17}), and 
\begin{equation}
\label{bos12}Z^{m_1\ldots m_{2q-2k}}=\int\limits_{W\left( t\right)
}dX^{m_1}\cdots dX^{m2q-2k}\cdot \frac{\left( dA\right) ^k}{k!}\quad = 
\end{equation}
\begin{equation}
\label{bos121}=\int\limits_{W\left( t\right) }d^{2q}\sigma \,\frac{\epsilon
^{0\mu _1\ldots \mu _{2q-2k}\nu _1\ldots \nu _{2k}}}{\left( 2q-2k\right)
!\,2^k\cdot k!}\cdot \partial _{\mu _1}X^{m_1}\cdots \partial _{\mu
_{2q-2k}}X^{m_{2q-2k}}\cdot F_{\nu _1\nu _2}\cdots F_{\nu _{2k-1}\nu
_{2k}}\quad . 
\end{equation}
Moreover, the relation between $q$ and $S\left( 2q\right) $ is given in
Table \ref{Tabelle1}. Note that the integrand of the charge $Y^m$ is closed
on the physical trajectories, see (\ref{an171}).

At last, from (\ref{an26}) and (\ref{an28}) we learn that 
\begin{equation}
\label{bos13}\left[ Q_\alpha ,P_m\right] =0\quad ,\quad \left[
P_m,P_n\right] =0\quad . 
\end{equation}
To avoid confusion we emphasize that in (\ref{bos10}) the bracket $\left\{
Q_\alpha ,Q_\beta \right\} $ denotes a graded {\bf Poisson}-bracket between
two Grassmann-odd quantities, but in (\ref{bos13}) we have chosen a square
bracket to denote the {\bf Poisson} bracket between quantities of which at
least one of them is Grassmann-even.

\subsection{Interpretation of the central charges}

Let us try to interprete the structure of the charges $Z^{m_1\ldots
m_{2q-2k}}$ in (\ref{bos12}). Let us fix $q$ and first of all look at the
extreme values of $k$, i.e. $k=0$ and $k=q$. For $k=0$ we find 
\begin{equation}
\label{bos14}Z^{m_1\ldots m_{2q}}=\int\limits_{W\left( t\right)
}dX^{m_1}\cdots dX^{m_{2q}}\quad ; 
\end{equation}
from (\ref{gract40}) we see that this is just the integral over the
topological current $j_T^{0m_1\cdots m_{2q}}$, i.e. the topological charge 
\begin{equation}
\label{bos15}Z^{m_1\ldots m_{2q}}=T^{m_1\ldots m_{2q}} 
\end{equation}
from (\ref{gract401}). This charge will not be defined if the brane $W\left(
t\right) $ is infinitely extended in one of the spatial directions $X^{m_i}$
occuring in $T^{m_1\ldots m_{2q}}$. On the other hand, if all spacetime
directions occuring in $T^{m_1\ldots m_{2q}}$ are compact, but the brane is
not wrapped around {\bf all} of them then this charge will be zero. It will
be non-zero {\bf only} if the brane wrappes around all these compact
dimensions; consider, for example, a compact $U\left( 1\right) $-factor in
the spacetime, which may be taken as direction $m=1$, and a closed string
that is wrapped around this $n$ times \cite{Azca} (the string is not a IIA
brane, of course, but that does not affect the discussion here); then $T^1$
is proportional to $2n\pi $, where $n$ is an integer. On the other hand, if
the string is closed in a flat spacetime, then $T^1=0$.

For $k=q$ we find that 
\begin{equation}
\label{bos16}Z=\frac 1{q!}\int\limits_{W\left( t\right) }\left( dA\right)
^q\quad . 
\end{equation}
This can be given a simple interpretation in the case of $q=1$, $p=2q$, i.e.
the D-$2$-brane: In this case 
\begin{equation}
\label{bos17}Z=\int\limits_{W\left( t\right) }dA 
\end{equation}
is just the flux of the field strength $F$ of the gauge potential through
the brane $W\left( t\right) $. The worldvolume is now to be regarded as a $%
U\left( 1\right) $-bundle $P\left( W,U\left( 1\right) \right) $. If the
section $W\left( t\right) $ is infinitely extended then $Z$ will vanish
provided that the gauge field vanishes sufficiently fast at infinity, and
the bundle is trivial. If, however, $W\left( t\right) $ describes a $S^2$,
say, then we have the possibility that the gauge potential is no longer
defined globally on $S^2$; if two gauge patches are necessary to cover $S^2$
then the flux integral (\ref{bos17}) yields 
\begin{equation}
\label{bos18}Z=4\pi g\quad ,\quad 2g\in {\bf Z}\quad , 
\end{equation}
where $g$ now is the charge of a {\it Dirac monopole} of the gauge field
sitting ''in the centre of $S^2$'', and the quantization condition $2g\in 
{\bf Z}$ comes from the requirement that the transition function between the
two gauge patches be unique, see for example \cite{Nakahara}. It is not
clear to us whether this interpretation extends to all possible values of $q$%
; we might conjecture that the $U\left( 1\right) $-bundle can always be
non-trivial, in which case similar arguments apply to (\ref{bos16}), since
then we must cover $W\left( t\right) $ by more than one gauge patch, which
should yield analogous results.

As for the values $1\le k\le q$ we see that the currents $dX^{m_1}\cdots
dX^{m2q-2k}$ in (\ref{bos12}) now probe whether the brane has subcycles of
dimension $\left( 2q-2k\right) $ embedded in it that wrap around $\left(
2q-2k\right) $ compact dimensions of the spacetime. Only in this case the
charges $Z^{m_1\ldots m_{2q-2k}}$ will be non-vanishing. Furthermore we see
that the $U\left( 1\right) $-bundle defined by the gauge field must be
non-trivial in order to having a non-vanishing charge. To see this we can
choose a static gauge $\sigma ^\mu =X^\mu $, then from (\ref{bos121}) we
have that 
\begin{equation}
\label{bos19}Z^{m_1\ldots m_{2q-2k}}=\int\limits_{W\left( t\right)
}d^{2q}\sigma \,\frac{\epsilon ^{0m_1\ldots m_{2q-2k}\nu _1\ldots \nu _{2k}}%
}{\left( 2q-2k\right) !\cdot k!}\cdot \partial _{\nu _1}A_{\nu _2}\cdots
\partial _{\nu _{2k-1}}A_{\nu _{2k}}\quad . 
\end{equation}
(This static gauge will be allowed at least on a certain coordinate patch on
the worldvolume; in this case we have to sum over contributions from the
different patches). We see that similar considerations concerning the
non-triviality of the $U\left( 1\right) $-bundle should apply here. In
particular, (\ref{bos19}) will vanish if the bundle is trivial, since in
this case the gauge potential $A_\mu $ is globally defined, and then (\ref
{bos19}) yields a surface term. A tentative interpretation of the charges (%
\ref{bos12}) therefore would be that they measure the coupling of compact
spacetime dimensions the brane or some directions of the brane wrap around
to non-trivial gauge field configurations on the brane.

We have not found an easy interpretation for $Y^m$; from its structure we
see that the $dX^m$-factor together with the fact that $*\Lambda ^{gauge}$
is closed on the physical trajectories will make this charge non-vanishing
only when the brane contains a $1$-cycle wrapping around a compact spacetime
dimension, e.g. a $S^1$-factor. This charge then describes the coupling of
the canonical gauge field momentum to this particular topological
configuration.

We finally present the modified charge algebra in the case of the D-$2$%
-brane with worldvolume ${\bf R}\times S^2$, since this allows for an easy
interpretation, as we have seen above:%
$$
\left\{ Q_\alpha ,Q_\beta \right\} =2\left( C\Gamma ^m\right) _{\alpha \beta
}\cdot P_m\;-\;2i\left( C\Gamma _{11}\Gamma _m\right) _{\alpha \beta }\cdot
Y^m\;- 
$$
\begin{equation}
\label{bos20}-\;i\left( C\Gamma _{m_2m_1}\right) _{\alpha \beta }\cdot
T^{m_1m_2}\;-\;2i\left( C\Gamma _{11}\right) _{\alpha \beta }\cdot 4\pi
g\quad , 
\end{equation}
where $T^{m_1m_2}$ probes the presence of compact dimensions in spacetime
the brane wraps around, and $g$ is the quantized charge of a possible Dirac
monopole resulting from the gauge field.

\end{document}